\documentclass[%
superscriptaddress,
reprint,
showpacs,
 amsmath,amssymb,
 aps,
pra,
]{revtex4-1}

\usepackage{color}
\usepackage{graphicx}
\usepackage{dcolumn}
\usepackage{bm}
\usepackage{amssymb}

\newcommand{\threejm}[6]{ \left(\begin{array}{ccc} #1 & #3 & #5\\
                                              #2 & #4 & #6
                                \end{array}
                          \right)}

\newcommand{\ninej}[9]{\left\{\begin{array}{ccc}
  #1 & #2 & #3\\
  #4 & #5 & #6\\
  #7 & #8 & #9
\end{array}
\right\}}

\begin{document}

\title{
Spin coherence and optical properties of alkali-metal atoms  in solid parahydrogen
}

\author{Sunil Upadhyay}
\affiliation{Department of Physics, University of Nevada, Reno NV 89557, USA}
\author{Ugne Dargyte}
\affiliation{Department of Physics, University of Nevada, Reno NV 89557, USA}
\author{Vsevolod D. Dergachev}
\affiliation{Department of Chemistry, University of Nevada, Reno NV 89557, USA}
\author{Robert P. Prater}
\affiliation{Department of Physics, University of Nevada, Reno NV 89557, USA}
\author{Sergey A. Varganov}
\affiliation{Department of Chemistry, University of Nevada, Reno NV 89557, USA}
\author{Timur V. Tscherbul}
\affiliation{Department of Physics, University of Nevada, Reno NV 89557, USA}
\author{David Patterson}
\affiliation{Broida Hall, University of California, Santa Barbara, Santa Barbara, California 93106, USA}
\author{Jonathan D. Weinstein}
\email{weinstein@physics.unr.edu}
\homepage{http://www.weinsteinlab.org}
\affiliation{Department of Physics, University of Nevada, Reno NV 89557, USA}


\begin{abstract}
We present a joint experimental and theoretical study of spin coherence properties of  $^{39}$K, $^{85}$Rb, $^{87}$Rb, and $^{133}$Cs atoms trapped in a solid parahydrogen matrix.
We use optical pumping to prepare the spin states of the implanted atoms and circular dichroism to measure their spin states. 
Optical pumping signals show order-of-magnitude differences depending on both matrix growth conditions and atomic species.
We measure the ensemble transverse relaxation times (T$_2^*$) of the spin states of the alkali-metal atoms.
Different alkali species exhibit dramatically different T$_2^*$ times, ranging from sub-microsecond coherence times for high $m_F$ states of $^{87}$Rb, to $\sim 10^2$ microseconds for $^{39}$K. These are the longest ensemble T$_2^*$ times reported for an electron spin system at high densities ($n \gtrsim 10^{16}$~cm$^{-3}$).
To interpret these observations, we develop a theory of inhomogenous broadening of hyperfine transitions of $^2$S atoms in weakly-interacting solid matrices. Our calculated ensemble transverse relaxation times agree well with experiment, and suggest ways to longer coherence times in future work.
\end{abstract}

\maketitle

\section{Introduction}

Addressable solid-state electron spin systems are of interest for many physics applications, including  quantum computing and quantum information \cite{buch2013spin, tyryshkin2012electron, 2009_NV_isotope, robledo2011high, childress2013diamond, PhysRevLett.102.210502}, magnetometery \cite{PhysRevB.80.115202, taylor2008high, bauch2018ultralong}, nanoscale magnetic resonance imaging \cite{PhysRevX.5.011001, staudacher2013nuclear, mamin2013nanoscale, sushkov2014magnetic}, and tests of fundamental physics \cite{arndt1993can, kinoshita1994optical, vutha2018oriented, PhysRevA.98.032513}.

Atoms trapped in inert matrices --- such as hydrogen or noble-gas solids --- are promising for these applications. The transparent matrix allows for optical pumping and probing of the electron spin state of the implanted atom, and the weak interaction of the trapped atom with the host matrix should only minimally perturb the atomic properties. The hope is to combine the high densities of solid-state electron spin systems with the (marginally perturbed) excellent properties of gas-phase atoms.

Cesium atoms in the bcc phase of solid helium (at pressures of $\sim 26$ bar and temperatures of $\sim 1.5$~K) exhibit good optical pumping and readout of spin states and excellent ensemble spin coherence times, but to date have been limited to  low cesium densities ($\lesssim 10^{9}$~cm$^{-3}$) \cite{PhysRevA.60.3867, Weis1996, moroshkin2006spectroscopy}.
On the other hand, atoms can be trapped in argon and neon matrices at high densities ($\gtrsim 10^{17}$~cm$^{-3}$) \cite{PhysRev.137.A490, PhysRevLett.107.093001, PhysRevA.99.022505}, but to date optical pumping and readout of the electron spin state has been significantly less efficient than the best solid state spin systems \cite{PhysRev.166.207, kanagin2013optical}.

Parahydrogen is a promising cryogenic host matrix \cite{Momose1998} which combines the respective advantages of solid argon and solid helium.
Previously it was demonstrated that the spin state of rubidium in solid parahydrogen could be optically pumped and probed more efficiently than in solid argon \cite{upadhyay2016longitudinal}. Moreover, demonstrated ensemble electron spin coherence times for Rb atoms in solid parahydrogen are longer than any other solid-state system capable of comparable electron spin density \cite{PhysRevB.100.024106}.

In this work, we compare the optical pumping properties and ensemble transverse spin relaxation time (T$_2^*$) for potassium, rubidium, and cesium in solid H$_2$. 
The dramatic differences between these alkali-atom species reveal the underlying physical mechanisms affecting optical pumping and spin coherence times.

We further develop a first-principles theoretical model to describe the coherence properties of matrix-isolated alkali-metal atoms, which shows that the measured T$_2^*$ times are due to the anisotropic hyperfine interaction of the atoms with the host matrix. Our theoretical results are in good agreement with experiment, opening up the possibility of systematic \textit{ab initio} modeling of coherence properties of atomic and molecular guest species in inert matrices.

\section{Experiment}

\begin{figure}[ht]
    \begin{center}
    \includegraphics[width=\linewidth]{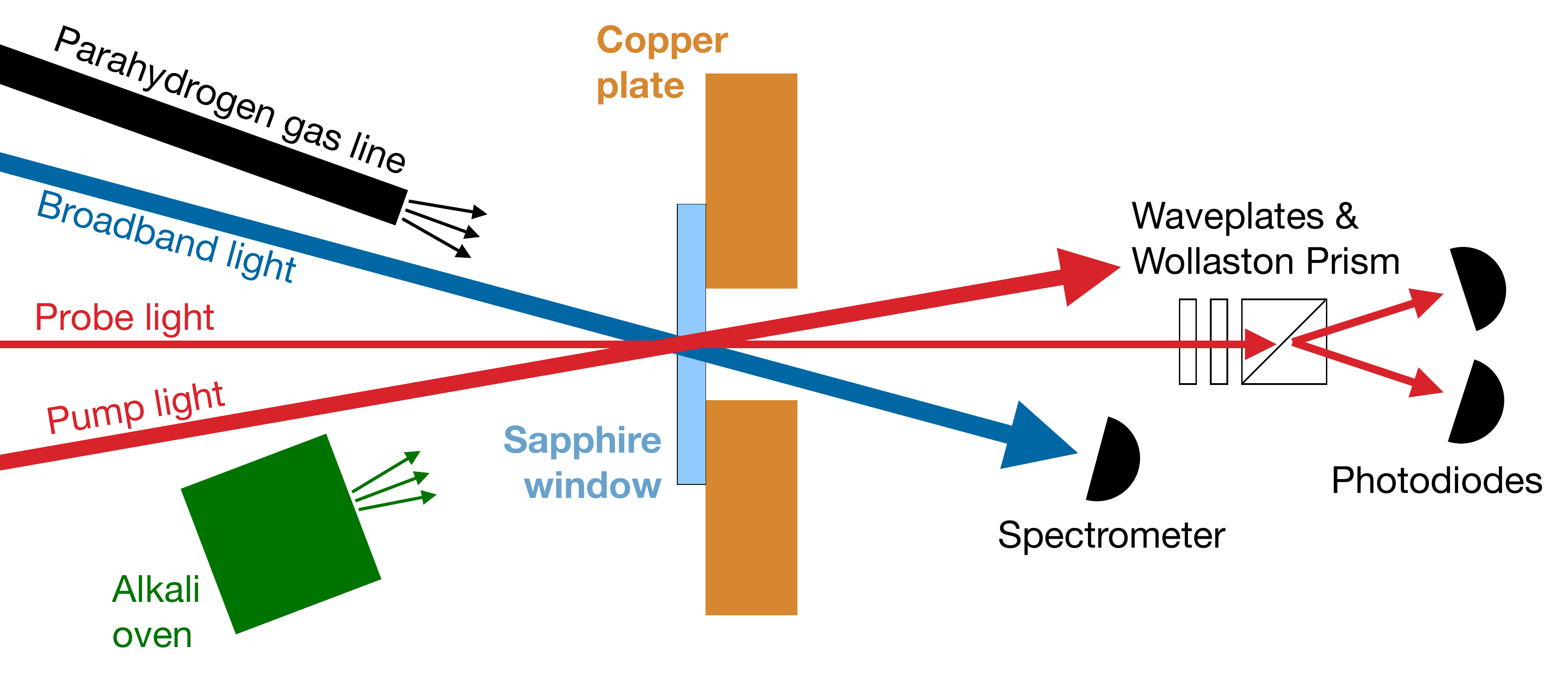}
    \caption{ 
	Schematic of apparatus. The copper plate and parahydrogen gas line are cooled by a closed-cycle pulse tube refrigerator.
	In our experiments, the pump and probe light are of the same frequency, and generated from the same laser. The vacuum chamber, its windows, and other optics are omitted for simplicity.
    \label{fig:ExpSchematic}}
    \end{center}
\end{figure}

The apparatus is as described in references \cite{upadhyay2016longitudinal, HartzellThesis, KanaginThesis}; the key components are shown in Fig. \ref{fig:ExpSchematic}.
Parahydrogen and alkali atoms (from high-purity, natural isotopic abundance samples) are co-deposited onto a cryogenically-cooled sapphire substrate in vacuum. Before deposition, normal hydrogen is converted to  parahydrogen using a cryogenic catalyst \cite{upadhyay2016longitudinal, jcp.108.4237}. In our current apparatus the remaining orthohydrogen fraction can be varied from $3 \times 10^{-5}$ to $1 \times 10^{-2}$. After deposition, the atoms are optically pumped and probed with both broadband and laser light at near-normal angles of incidence to the surface. A homogeneous magnetic bias field is applied to the crystal, and RF magnetic fields can be applied perpendicular to the bias field.

\section{Optical absorption spectra}
\label{sec:OpticalSpectra}

\begin{figure}[ht]
    \begin{center}
    \includegraphics[width=\linewidth]{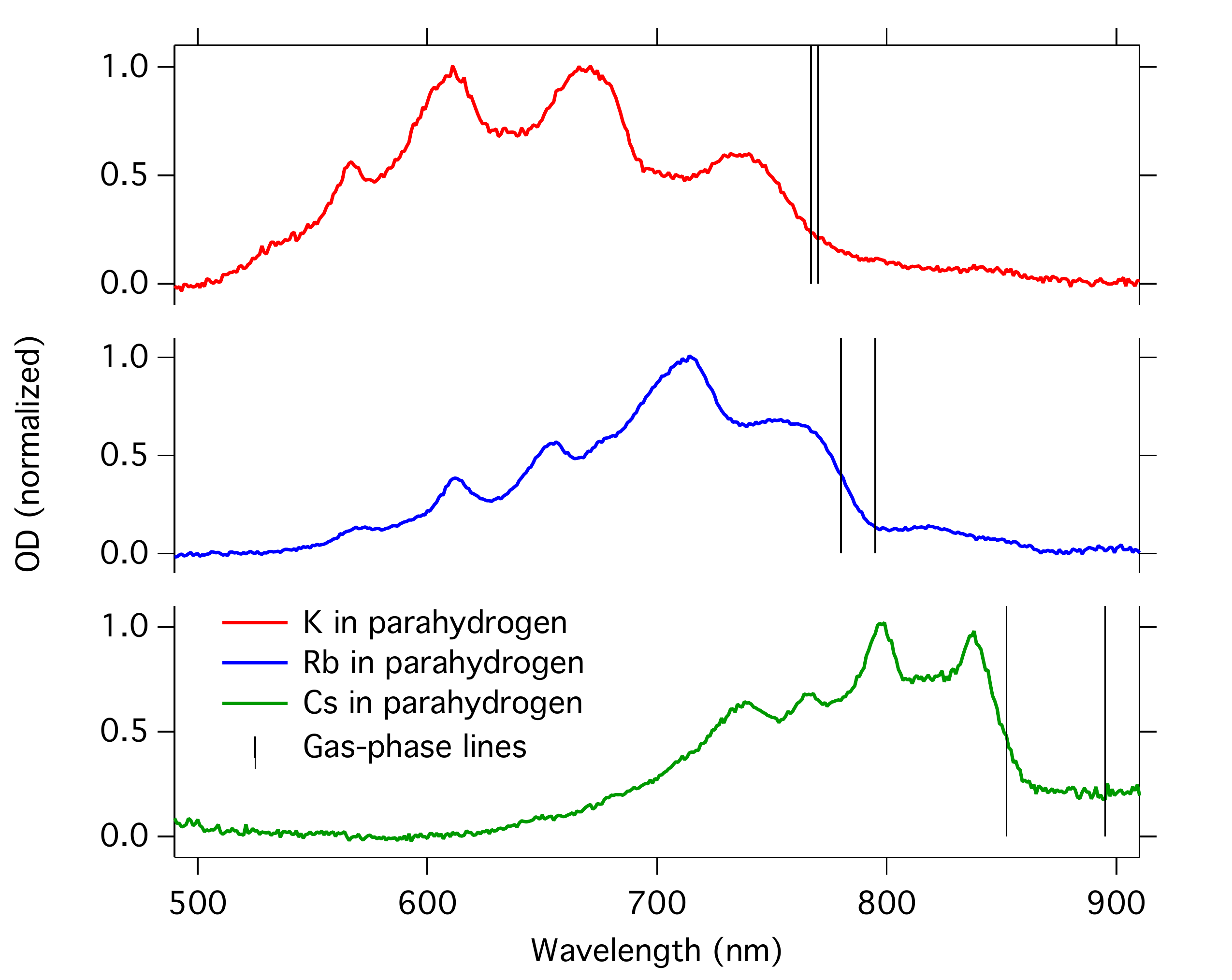}
    \caption{ 
Optical spectra of alkali-doped parahydrogen crystals. The spectra are normalized as discussed in the text. In each spectra, the frequencies of the gas-phase atom transitions \cite{NISTAtomicBasic} are shown for comparison.
\label{fig:opticalspectrum}
    }
    \end{center}
\end{figure}

Sample spectra of K, Rb, and Cs are shown in Fig.~\ref{fig:opticalspectrum}. The transmission $T$ of the crystal is determined by comparing a spectrum of the light transmitted through the apparatus --- as measured by a fiber-coupled grating spectrometer --- before and after crystal deposition. The optical depth (OD) is determined from $T \equiv e^{-\mathrm{OD}}$. For ease of comparing spectra 
the baseline of the spectra have been shifted so that the off-resonance OD~$=0$; the amplitudes have been normalized so that the peak OD~$= 1$. 

All spectra shown were taken at the 3 Kelvin 
base temperature of the cryostat. We note that thermal annealing of the crystal at temperatures of up to  4.2~K  and times up to 24 hours 
causes negligible changes in the absorption spectrum.

In all spectra, we see large spectral shifts, large broadenings, and the splitting of the $s \rightarrow p$ transition into multiple lines.
Similar behavior was observed for alkali atoms in noble gas matrices and superfluid helium \cite{PhysRev.137.A490, PhysRevA.60.3867, moroshkin2006spectroscopy, gerhardt2012excitation, PhysRevLett.71.1035}.

\subsection{Optical annealing}

The spectra of the implanted alkali atoms --- if grown in the absence of light --- are significantly affected by the application of broadband light to the crystal. This phenomenon, which we call ``optical annealing'', has been previously reported for Rb atoms \cite{upadhyay2016longitudinal}. 
Similar effects were observed for Cs and K. 
%
Typically during optical annealing the number of spectral peaks is reduced, and the optical depth of the remaining peaks increases correspondingly.
As far as we know, these changes are irreversible; in our observations we have not observed the spectrum returning to its original form, even over timescales of weeks.
We attribute the spectral changes to the reconfiguration of trapping sites due to optical excitiation.

The data shown in Fig. \ref{fig:opticalspectrum} is after optical annealing.
We have not studied optical pumping of atoms prior to optical annealing (nor have we studied the spectral peaks that disappear in the process), as we expect those sites not to  be stable under optical excitation.
For the remainder of this paper, we only discuss the properties of samples in this state reached after optical annealing.

\subsection{Bleaching and broadening mechanisms}
\label{sec:Bleaching}

Much as the optical spectrum is changed by the application of broadband light, we observe that it is also altered by the application of narrowband light.

\begin{figure}[ht]
    \begin{center}
    \includegraphics[width=\linewidth]{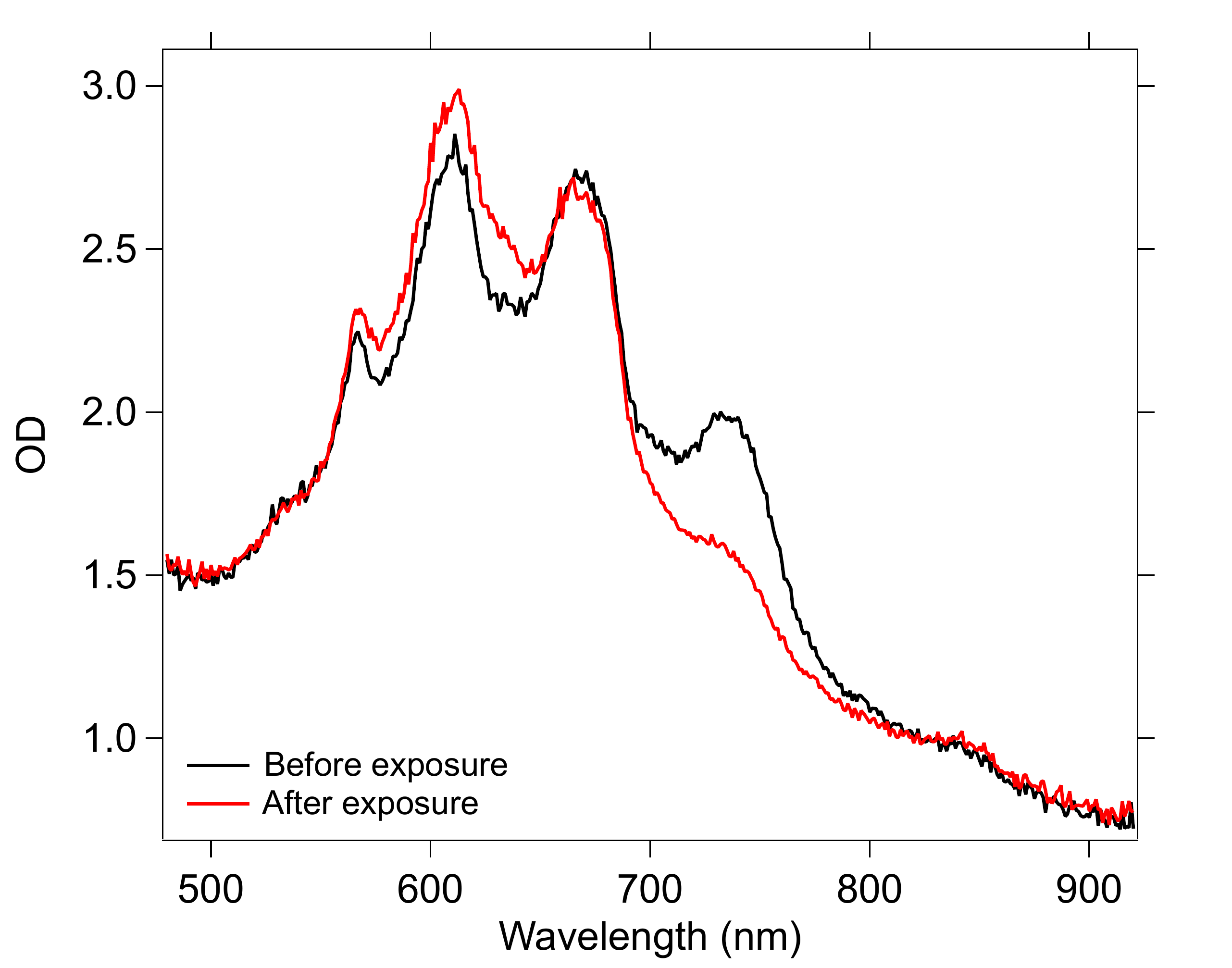}
    \caption{ 
Spectra of potassium-doped parahydrogen showing bleaching. Spectra are taken before and after illuminating the matrix with laser light at 757~nm. The light causes a significant reduction in the absorption of the peak it is on resonance with.
    \label{fig:bleaching}
    }
    \end{center}
\end{figure}

For potassium atoms trapped in parahydrogen, we see ``bleaching'' effects due to the application of narrowband light, as seen in Fig. \ref{fig:bleaching}. We attribute the changes in the spectrum to changes in the trapping sites induced by the light, similar to what occurs during  optical annealing with broadband light.

The changes to the spectrum indicate that broadening is homogenous within each peak: application of light with a linewidth $\ll$ the absorption linewidth effectively bleaches the entire line. The changes also indicate that the different lines originate from different trapping sites, as absorption at other frequencies is not diminished. In fact, absorption at 610~nm increases, indicating that during bleaching the trapping sites that give rise to absorption at 735~nm are changed into trapping sites that absorb at a different frequency.

Similar bleaching effects were observed for Rb atoms trapped in solid argon. We note in argon, $\sim 10^1$ photon scattering events would cause reconfiguration of the trapping sites \cite{kanagin2013optical}. Alkali atoms in parahydrogen are significantly more resistant to bleaching. From the atomic density, the intensity of light, and the timescale of bleaching, we estimate potassium 
absorbs on the order of $10^4$ photons before bleaching.

Such bleaching effects can be problematic for use of these matrix-trapped atoms for applications. For Rb atoms in argon we found that application of light at other wavelengths would reverse the bleaching effects and return the trapping sites to their ``unbleached'' states \cite{kanagin2013optical}. We have not yet demonstrated similar unbleaching with alkali atoms in parahydrogen, it is not yet known if this is possible.

%
%

\subsection{Effects of crystal growth conditions}
\label{sec:crystal_growth_optical}

The spectra of alkali atoms in parahydrogen can vary significantly with crystal growth conditions.

We did not observe a significant dependence of the spectra on alkali density or orthohydrogen density over the ranges we explored.
We saw no noticeable change with ortho fraction over the  range from $4\times 10^{-5}$ to $3\times 10^{-3}$. 
Similarly, the spectra show only minor changes with Rb atom density from $1 \times 10^{17}$~cm$^{-3}$ to $1 \times 10^{18}$~cm$^{-3}$.
%
%
%
%
However, the spectra do depend sensitively on the substrate temperature  at the time of matrix growth.

\begin{figure}[ht]
    \begin{center}
    \includegraphics[width=\linewidth]{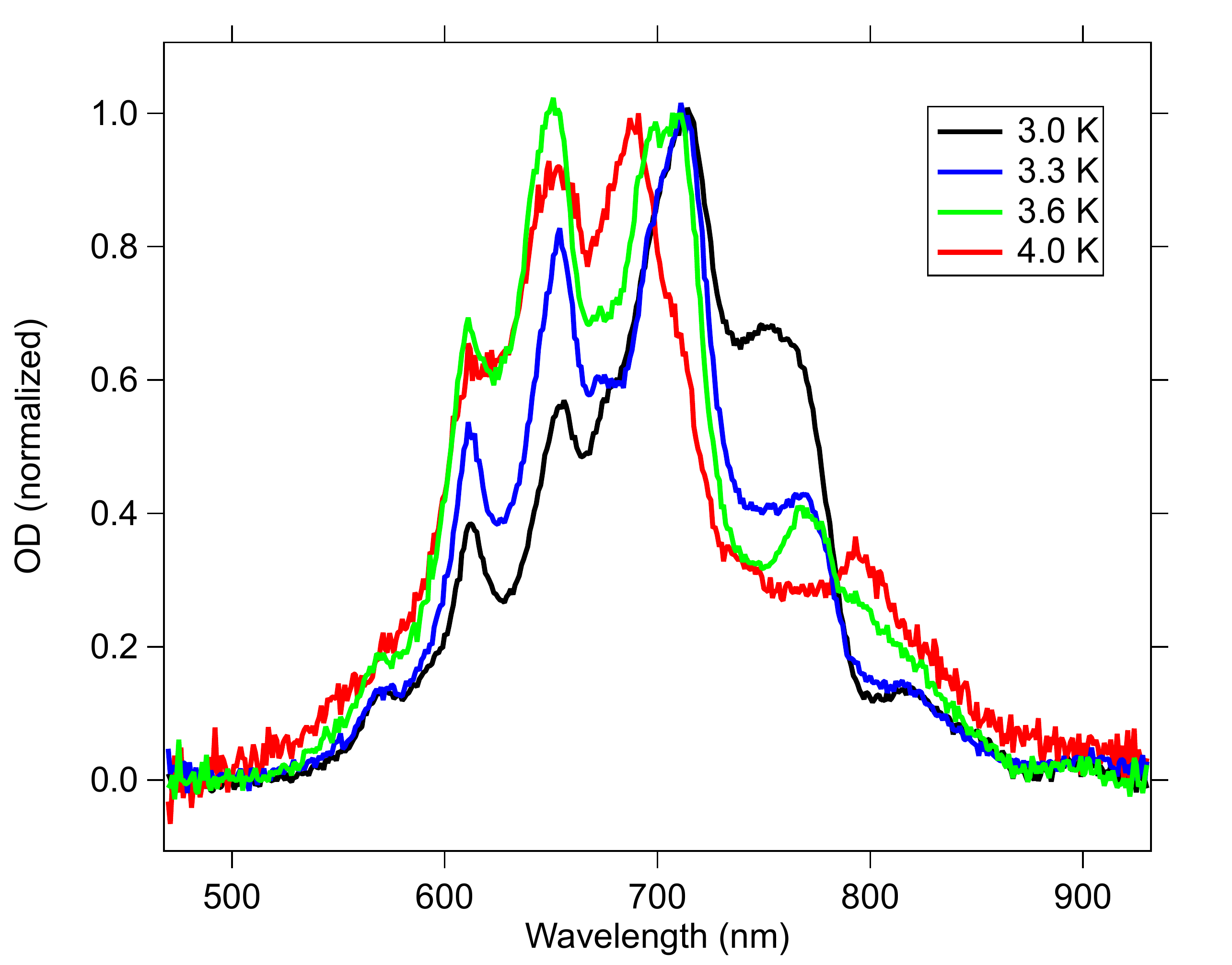}
    \caption{ 
Optical spectra of Rb-doped parahydrogen crystals, grown at different substrate temperatures, as labeled. The spectra are normalized in the same manner as Fig. \ref{fig:opticalspectrum}.
%
%
%
%
%
%
    \label{fig:Rb_optical_spectrum_vs_growth_T}
    }
    \end{center}
\end{figure}

Figure \ref{fig:Rb_optical_spectrum_vs_growth_T} shows the optical spectra of Rb-doped parahydrogen crystals grown at different substrate temperatures. The temperatures specified in the figure --- and throughout this paper --- are of the copper plate upon which the sapphire window is mounted. The temperature of the crystal itself is necessarily warmer than this substrate. Measurements of the front surface of the window (onto which the parahydrogen matrix is grown) indicate that its temperature is within 0.4~K 
 of the copper plate temperature.
While the crystals in Fig. \ref{fig:Rb_optical_spectrum_vs_growth_T} were grown at different substrate temperatures, the spectra shown were measured under identical conditions at our base temperature of 3 K. 
%
The crystals of Fig.~\ref{fig:Rb_optical_spectrum_vs_growth_T} have Rb densities of $1 \times 10^{17}$~cm$^{-3}$, with variations within $\pm 15$\%, and ortho fractions of $3 \times 10^{-5}$, with variations of $\pm 25$\%. We believe the spectral differences are primarily due to the substrate temperature. 

As the substrate temperature increases, the blue-shifted peaks become larger in amplitude, and the red-most peak becomes smaller and shifts. Qualitatively similar behavior was observed for Cs atoms. 
(Potassium-doped matrices were only grown at the base temperature.)

Subsequent annealing at temperatures up to 4~K for durations up to $\sim 10$ hours can change the background scattering from the crystal (depending on crystal conditions, annealing has been observed to either increase or decrease background scattering). However, annealing has little observable effect on the alkali atom absorption peaks.


The optical spectrum is also affected by the matrix growth rate. Rubidium-doped crystals grown at our base temperature (3 K) with low hydrogen deposition rates ($\sim 1$~$\mu$m per minute) have optical spectra similar to samples grown at normal deposition rates ($\sim 3$~$\mu$m per minute) and higher substrate temperatures (similar to the 3.3 K spectrum shown in figure \ref{fig:Rb_optical_spectrum_vs_growth_T}). 
However, we have not explored flow rates as comprehensively as substrate temperatures.

As discussed in section \ref{sec:SpinPolSignal}, these changes in the optical spectrum have significant consequences for our ability to optically pump and measure the spin states of the alkali atoms.

\section{Spin polarization signal}
\label{sec:SpinPolSignal}

We optically pump the implanted atoms using right-hand circular (RHC) laser light, as shown in Fig. \ref{fig:ExpSchematic}. We monitor the spin polarization produced using a linearly-polarized probe beam at the same frequency. After passing through the sample, the probe beam is sent through waveplates and a Wollaston prism to separate it into its RHC and LHC components, which are measured on two photodetectors. Differential absorption between the RHC and LHC components (circular dichorism) indicates spin polarization. 
Due to the large  broadening of the optical spectrum, the different isotopes and their hyperfine levels cannot be optically distinguished, and the spin polarization signal measured for each species is an average of the naturally occurring isotopes.

To quantify the spin polarization obtained, we measure the ratio of RHC and LHC signals on the two photodiodes and normalize the ratio to a level of 1 before optical pumping. The ratio changes after optical pumping. To ensure that the change is not due to  systematic effects, it is measured both with an applied longitudinal magnetic field and with a transverse field (the ambient earth magnetic field); the latter prevents the accumulation of spin polarization during optical pumping. With a transverse field, the change in the ratio due to optical pumping is typically very small compared to the longitudinal field, as expected \cite{upadhyay2016longitudinal}. To calculate the polarization signal amplitude, one ratio is subtracted from the other.
This is the polarization signal $P$ shown below in Figures \ref{fig:CsPolSpectrum}, \ref{fig:PolVsGrowthT}, and \ref{fig:PolVsB}.

We relate this signal to atomic properties by the following model. Optical pumping changes the hyperfine and spin state of the implanted atoms. This changes the atoms' cross-section for scattering RHC and LHC light. We quantify the change with a single parameter $\Delta$, and model the cross-section changes as $\sigma_\mathrm{RHC} = \sigma_0 (1-\Delta)$, and $\sigma_\mathrm{LHC}= \sigma_0 (1+\Delta)$, consistent with our observations \cite{upadhyay2016longitudinal}. Before pumping (or after pumping with a transverse magnetic field) we assume $\Delta=0$, giving identical optical depths  for both polarizations of light; we refer to this optical depth as OD$_0$. Thus, when we measure the ratio $R$ of transmissions of LHC and RHC light, we obtain $R=e^{-2 \Delta \cdot \mathrm{OD}_0}$. The polarization signal $P$ we measure is then $P=1-R$. In the limit $P\ll 1$,  $P=2 \Delta \cdot \mathrm{OD}_0$.
For vapor phase  atoms, it is possible to obtain $\Delta \rightarrow 1$, as the atoms can be pumped into a spin state that is dark to one of the circular polarizations of light. As presented below, the largest values of $\Delta$ observed for alkali atoms in parahydrogen are $\sim0.065$, significantly lower than vapor phase atoms. 
Whether this is due to limitations in optical pumping or optical detection is not known; we expect it is a combination of both.


\subsection{Wavelength dependence}
\label{sec:PolVsWavelenth}

\begin{figure}[ht]
    \begin{center}
    \includegraphics[width=\linewidth]{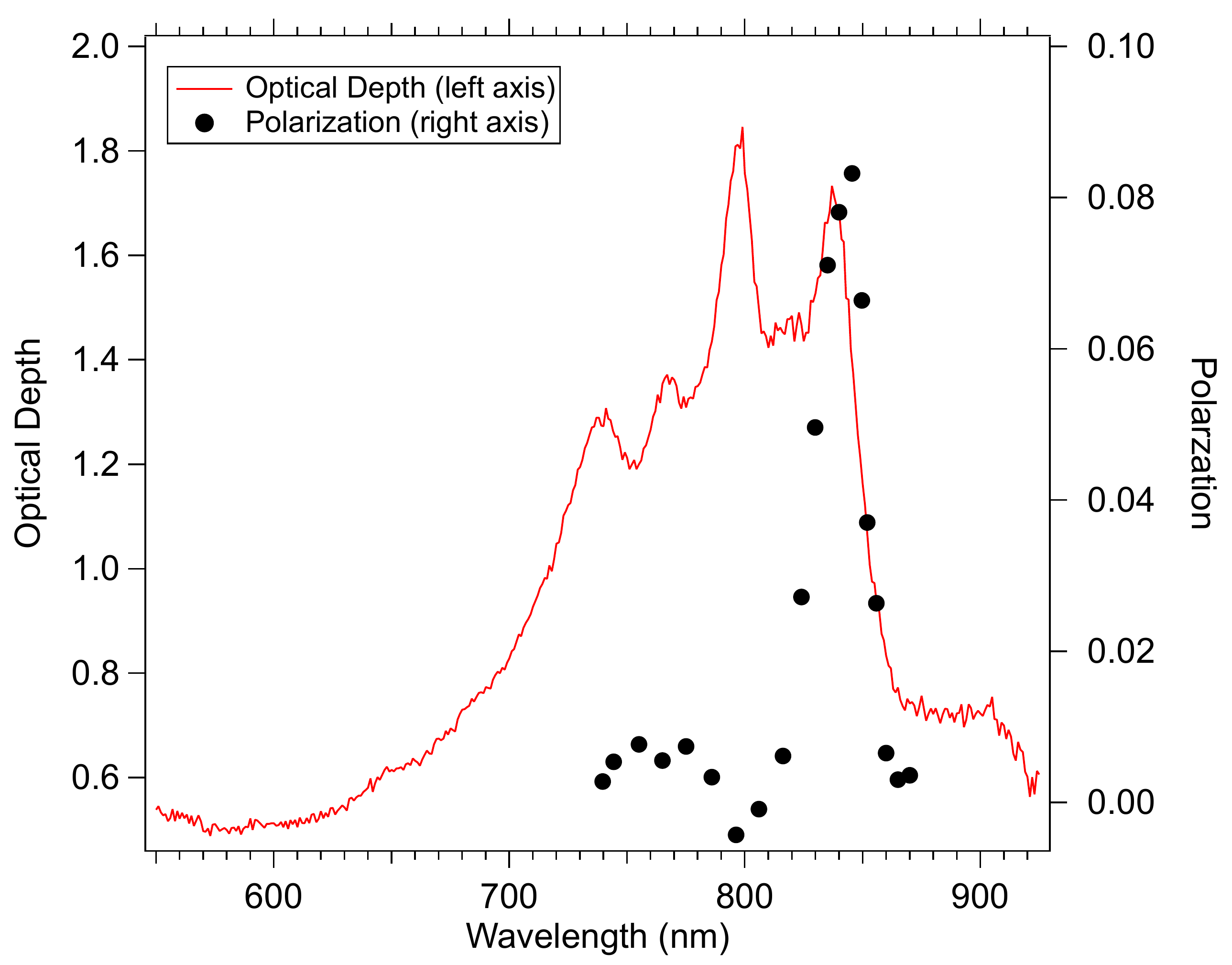}
    \caption{ 
Polarization signal amplitude, plotted alongside the optical spectra of a Cs-doped parahydrogen crystal. Crystal grown with a substrate temperature of 3~K, orthohydrogen fraction $3 \times 10^{-5}$, and Cs density $6 \times 10^{16}$~cm$^{-3}$. Polarization was measured with a 80~G on-axis bias field. 
    \label{fig:CsPolSpectrum}}
    \end{center}
\end{figure}

We examine the polarization signal as a function of the wavelength of the pump and probe  (the two wavelengths are identical in all data presented here). 
For the cesium data presented in Fig. \ref{fig:CsPolSpectrum}, typical pump and probe beams have waists of 200~$\mu$m and 125~$\mu$m respectively and intensities 5 $\times 10^{3}$ mW/cm$^{2}$ and 50 mW/cm$^{2}$ respectively. 
We note that while these intensities are  above the saturation intensity of a gas-phase alkali atom, they are far below the saturation intensity of alkali atoms in parahydrogen (due to the large spectral broadening of the optical transition).
Typical pump durations are $\sim$ 100 ms; the pumping rate is limited by laser intensity.
 %
%
While the optical spectrum shows multiple peaks, we only see a significant polarization signal when pumping and probing near the red-most peak.

For Rb polarization, we were not able to scan the entire Rb spectra, and were only able to cover the ranges 655nm, 680--700nm, and 730--810nm (limited by the light sources available to us). Over this range, we saw negligible polarization signal except in the region from 730--760~nm, with maximum signal near 750~nm. Similarly to Cs, the largest polarization signal was seen near the red-most peak (for crystals grown at low temperature).

For potassium, we likewise were unable to cover the entire spectral range, but were able to compare pumping and probing on the red-most line (at 735 nm) to the line at 660~nm. We unable to observe spin polarization at 660~nm, but observed a signal when slightly red-detuned from the red-most line.

This behavior is similar to what was previously observed for thermally-spin-polarized rubidium atoms in argon, which gave the strongest circular dichroism signal on the red-most line \cite{PhysRev.166.207}. We do not know whether this is due to similar physics or is simply a coincidence.

Surprisingly, for Rb spectra grown at elevated temperatures (as shown in Fig. \ref{fig:Rb_optical_spectrum_vs_growth_T}), the peak polarization response remains near 750~nm despite the nearly complete ``disappearance'' of that peak in the absorption spectrum. However, the size of the polarization signal decreases, as discussed below in section \ref{sec:PolVsGrowth}.

Much as the optical spectrum has little dependence on the ortho fraction or alkali density over the ranges we explored, we observed little change in the polarization signal.
For rubidium densities from $6 \times 10^{16}$ to $3 \times 10^{17}$~cm$^{-3}$ in matrices grown under similar conditions, we see no change in $\Delta$ to within $\pm 15\%$.
We note that at higher rubidium densities ($\gtrsim 10^{18}$~cm$^{-3}$) the polarization signal appeared to decrease, but we did not extensively explore this density region.
Increasing the ortho fraction from $5 \times 10^{-3}$ to $3 \times 10^{-2}$ resulted in a decrease in the Rb polarization signal of a factor of 2. However, the higher ortho fraction crystal was grown at a substrate temperature 0.16~K higher than the low ortho fraction crystal (due to the extra heat load on the cryostat from heating the ortho-para catalyst), and we suspect the majority of the difference in polarization signal is due to the substrate temperature change (as discussed below in section \ref{sec:PolVsGrowth}).
We did not investigate this behavior for Cs and K in a controlled manner.

\subsection{Effects of crystal growth conditions}
\label{sec:PolVsGrowth}

\begin{figure}[ht]
    \begin{center}
    \includegraphics[width=\linewidth]{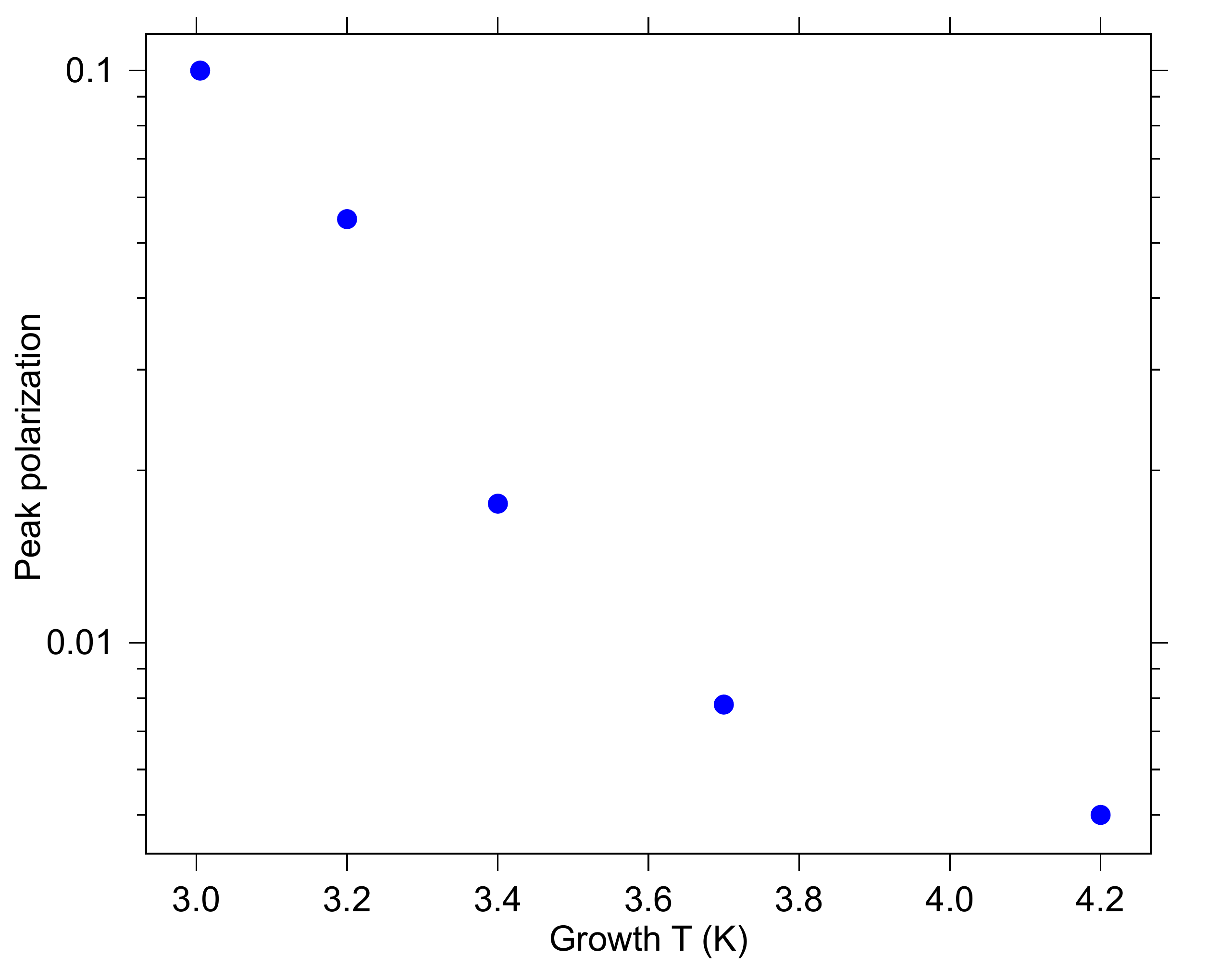}
    \caption{ 
Polarization signal amplitude for Cs-doped parahydrogen crystals of optical depth $\mathrm{OD}_0 \sim 1$ grown at different substrate temperatures. 
The optical depths of the different crystals differed by $\pm 25\%$, and their densities varied by $\pm 30\%$. As these variations are small compared to the polarization effects observed, we did not correct for them. The change in ortho fraction was small, and separate measurements indicate that ortho fraction has little effect on the size of the polarization signal.
    \label{fig:PolVsGrowthT}}
    \end{center}
\end{figure}

Because the crystal  growth temperature strongly affects the optical spectrum (as discussed in Section \ref{sec:crystal_growth_optical}), one might expect the polarization signal to be affected as well. 
This is indeed true: the size of the polarization signal varies  strongly with the temperature of the substrate during crystal growth. Fig. \ref{fig:PolVsGrowthT} shows this effect for the case of cesium atoms.

We note that the optical pumping data in Fig. \ref{fig:PolVsGrowthT} was all obtained at our base temperature, having cooled down the crystal after growth. Much as the optical spectrum maintains a ``memory'' of the temperature at which it was grown, so does the optical pumping and readout. 
Similar behavior was observed for Rb, with smaller polarization signals for crystals grown at elevated substrate temperatures.

For cesium, some data suggests that matrices grown at higher hydrogen deposition rates give larger polarization signals than samples grown at lower flow. This is consistent with the results of section \ref{sec:crystal_growth_optical}, indicating that higher hydrogen flow has a similar effect on the optical behavior as lower temperatures. The maximum flow rate is limited by our current ortho-para converter.

Based on these results and those of sections \ref{sec:OpticalSpectra} and \ref{sec:PolVsWavelenth}, we speculate that some trapping sites in the lattice are more favorable for optical pumping and readout. The different growth conditions change the fraction of atoms trapped in such favorable sites, which is reflected in both the optical spectrum and the polarization signal.

We note this data suggests it is very likely that significant improvements in the ability to optically pump and read out the spin states of alkali atoms in parahydrogen are possible with an apparatus capable of colder temperatures and faster parahydrogen deposition rates during crystal growth.

\subsection{Magnetic field dependence}
\label{sec:B_field_dependence_of_P}

\begin{figure}[ht]
    \begin{center}
    \includegraphics[width=\linewidth]{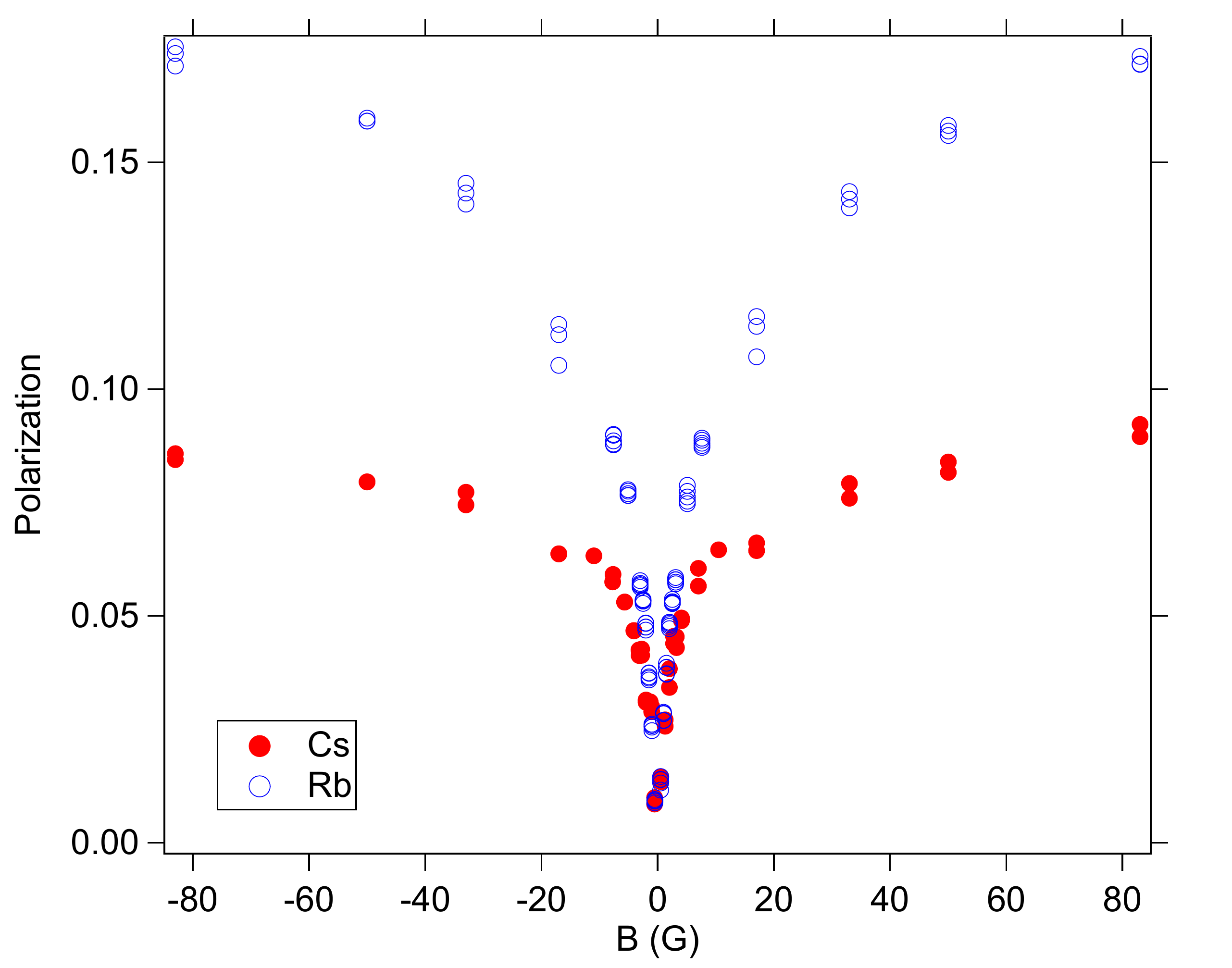}
    \caption{ 
Polarization signal $P$ for Cs and Rb as a function of the bias magnetic field. The bias field is roughly normal to the matrix surface and roughly parallel to the pump and probe beams.
The difference in the high-field value of $P$ is partially due to different growth conditions: the cesium-doped sample was grown at a higher substrate temperature; the OD of Rb  and Cs were 1.4 and 1.1 respectively. Both crystals exhibit the same qualitative behavior.
    \label{fig:PolVsB}}
    \end{center}
\end{figure}

As seen in figure \ref{fig:PolVsB}, the amplitude of the spin polarization signal has a strong dependence on the applied magnetic field. At fields $\ll 10$~Gauss, the optical polarization signal is quite small. The signal size increases with increasing magnetic field, and appears to saturate at fields $\gg 10$~G.

As discussed below in section \ref{sec:TheoryGeneric} (and touched upon previously in references \cite{upadhyay2016longitudinal, PhysRevA.60.3867}) we attribute these effects to coupling to the crystal field in our polycrystalline sample. At magnetic fields $\gg 10$~G, the Zeeman splitting is much larger that the coupling of the spin to the crystal field, and the $m$-levels are only slightly perturbed by the matrix. At low magnetic fields, the perturbation from the matrix mixes the $m$ eigenstates and interferes with the ability to optically control and probe the spin state with polarized light.

\subsection{Species dependence}
\label{subsec:OPspecies}

Potassium produces significantly smaller polarization signals than Rb and Cs-doped crystals produced and measured under similar conditions.

\begin{table}[ht]
\caption{ The optical spin-polarization signal $\Delta$, as defined in section \ref{sec:SpinPolSignal}, for the atomic species measured. All crystals had an optical depth of 1.1 at the pump/probe wavelength. The excited state fine structure splittings are from reference \cite{NISTAtomicBasic}.}
\begin{tabular}{l |  c| c || r}
Species & B (G) &  $\Delta$ & FS splitting (cm$^{-1}$)\\
\hline
K &  80  &  $4\times 10^{-3}$  & 57  \\ 
 Rb  & 33  & $5 \times 10^{-2}$ & 237  \\ 
 Cs  & 33 &  $4 \times 10^{-2}$ & 554  \\ 
\end{tabular}
  \label{tab:DeltaVsSpecies}
\end{table}


Table \ref{tab:DeltaVsSpecies} compares the spin polarization signals obtained for potassium, rubidium, and cesium. All crystals were grown on the identical cryogenic substrate setup and under similar growth conditions. The  potassium data is the largest polarization signal observed for potassium in our laboratory, and was measured before significant bleaching of the spot occurred (see section \ref{sec:Bleaching}). Larger signals were seen for rubidium and cesium crystals grown under different conditions (an improved window mount that was able to reach slightly colder temperatures, and higher parahydrogen flow rates). While the data was taken at different bias fields, rubidium and cesium polarizations do not have a significant dependence on the magnetic field over the range from 30 to 80 Gauss (as seen in figure \ref{fig:PolVsB}).

\subsubsection*{Interpretation}

As discussed below in section \ref{subsec:Abinitio}, all three species have similar ground-state interaction potentials with hydrogen.
Our interpretation is that the order-of-magnitude differences in polarization are due to the different fine-structure splittings of their excited states. 

Firstly, optical pumping and detection of spin polarization on the $s \rightarrow p$ transition in an alkali atom relies on the fine-structure coupling between orbital angular momentum ($L$) and spin ($S$). Inside the matrix, the excited $p$ orbital is coupled to the crystal field of the local trapping site, which (neglecting spin and fine structure), can split its threefold orbital degeneracy \cite{PhysRev.166.207}. If the coupling of $L$ to the crystal is large compared to the fine-structure splitting, it can potentially ``decouple'' $L$ and $S$ and impede the ability to both optically pump and detect the electron spin state \cite{PhysRev.166.207}. Hence, if the crystal field interaction is much larger than the fine structure splitting, we expect poor optical pumping and detection. 

Secondly, if the fine structure splitting of the excited state is not optically resolved, it will suppress the ability to optically detect spin polarization. However, we note that in the case of repopulation pumping,  optical  pumping would still be possible in this limit, as discussed in section \ref{sec:Depop_Repop}.

As expected from both these effects, for Rb and Cs --- with large fine-structure splittings --- we see large polarization signals; for potassium --- with a significantly smaller fine-structure splitting --- we see a smaller polarization signal.



\subsection{Nature of optical pumping}
\label{sec:Depop_Repop}

\begin{figure}[ht]
    \begin{center}
    \includegraphics[width=\linewidth]{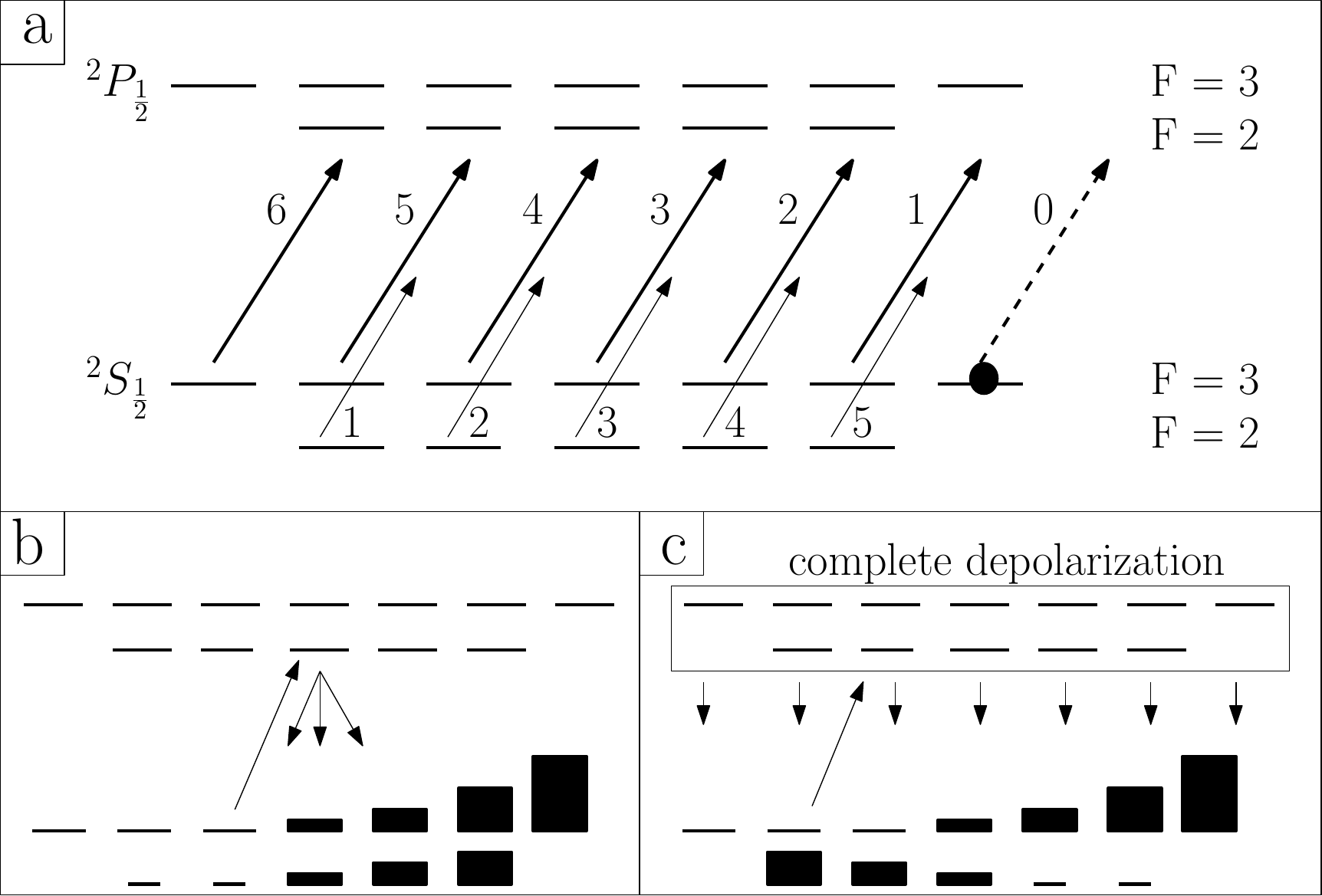}
    \caption{ 
Schematic of $^{85}$Rb optical pumping. Part (a) shows the relative line strengths of $\sigma^+$ transitions of the $^2S_{1/2} \rightarrow  \,   ^2P_{1/2}$  transition in the limit that the upper level hyperfine structure is unresolved \cite{metcalf1999laser}. Parts (b) and (c) show the expected populations in the case of repopulation and depopulation pumping, respectively, as discussed in the text. 
    \label{fig:Depol_vs_repol}}
    \end{center}
\end{figure}

Optical pumping of spin is  characterized as ``depopulation'' or ``repopulation'' pumping \cite{Happer72OptPumpReview}. In the depopulation limit, the excited state polarization state is completely randomized prior to decay to the ground state. In the repopulation limit the atomic polarization is conserved in the excited state. These two limits will lead to different spin state distributions, as shown in Fig. \ref{fig:Depol_vs_repol}.
For a free $^{85}$Rb atom driven on the $^2S_{1/2} \rightarrow  \,  ^2P_{1/2}$ transition, depopulation pumping will result in (semi-)dark states for both the $F=2$ and $F=3$ manifold. Repopulation pumping will produce a dark state in the $F=3$ but a bright state in $F=2$.

After optical pumping we can sweep an RF field to depolarize the ground state population. By monitoring the resulting changes in the optical signal we can distinguish between the two cases shown in Fig. \ref{fig:Depol_vs_repol}. As previously reported in reference \cite{PhysRevB.100.024106}, the polarization signal $P$ of $^{85}$Rb shifts in opposite directions for RF depolarization of the $F=2$ and $F=3$ levels. This indicates that the pumping is predominantly repopulation pumping. Similar behavior was seen for $^{87}$Rb, indicating that it also undergoes repopulation pumping. Cs and K were not measured in this manner.

For comparison, it was previously reported that optical pumping of the spin of cesium atoms in solid helium was predominantly repopulation pumping  \cite{PhysRevA.60.3867}, however rubidium atoms in solid helium underwent depopulation pumping \cite{PhysRevLett.88.123002}. 



\subsection{Comparison to argon}

In prior work, the spectra of alkali atoms trapped in argon matrices exhibited multiple absorption peaks, in groups of ``triplets'' \cite{PhysRev.137.A490, PhysRev.166.207, kanagin2013optical}. In those experiments each triplet was attributed to the crystal-field interaction splitting the three-fold degeneracy of the excited-state $p$ orbital. 

The bleaching results presented in section \ref{sec:Bleaching} suggest that the crystal-field splitting of the excited $p$ orbital in parahydrogen is too small to resolve. Our interpretation is that excited-state alkali atoms in parahydrogen experience a smaller crystal-field interaction than in argon. This may be the reason why the spin polarization signals seen for rubidium in parahydrogen are an order-of-magnitude  larger than the largest signals reported for rubidium in argon \cite{kanagin2013optical}.

\section{Longitudinal spin relaxation}

We can measure the longitudinal relaxation time, T$_1$, by observing the decay of the polarization $P$ over time. The T$_1$ of rubidium atoms in parahydrogen was previously reported in reference \cite{upadhyay2016longitudinal}.  It depends strongly on the orthohydrogen fraction in the crystal, with longer T$_1$ times at lower orthohydrogen fractions. T$_1$ is on the order of 1~s at ortho fractions $\lesssim 10^{-2}$ and magnetic fields $\gtrsim 10$~G.  At lower magnetic fields, T$_1$ is considerably shorter. We did not systematically measure the T$_1$ of Cs and K at high ortho fractions, but observed T$_1$ times on the order of 1~s at low ortho fractions. Cs showed a similar strong dependence on the magnetic field, with T$_1$ shorter at magnetic fields $\lesssim 10$~G, and saturating at higher fields.

What processes limit T$_1$ and whether longer times might be achieved is not understood at this time.
Our primary interest at present is in the ensemble transverse relaxation time T$_2^*$. As the measured T$_1 \gg \mathrm{T}_2^*$, longitudinal relaxation does not play a significant role in limiting T$_2^*$.





\section{Ensemble transverse spin relaxation}
\label{sec:T2*}

We measure the  ensemble transverse spin relaxation time (T$_2^*$) with  free-induction decay (FID) measurements, as well as other methods detailed in reference \cite{PhysRevB.100.024106}.
After optically pumping the spin state of the atoms, we apply a short RF pulse to induce Larmor precession and observe the resulting oscillations in the polarization signal. Because  different isotopes typically have different $g$-factors, we can frequency-select a single isotope with the RF pulse, allowing us to measure the FID signals of the different isotopes separately.

\begin{figure}[ht]
    \begin{center}
    \includegraphics[width=\linewidth]{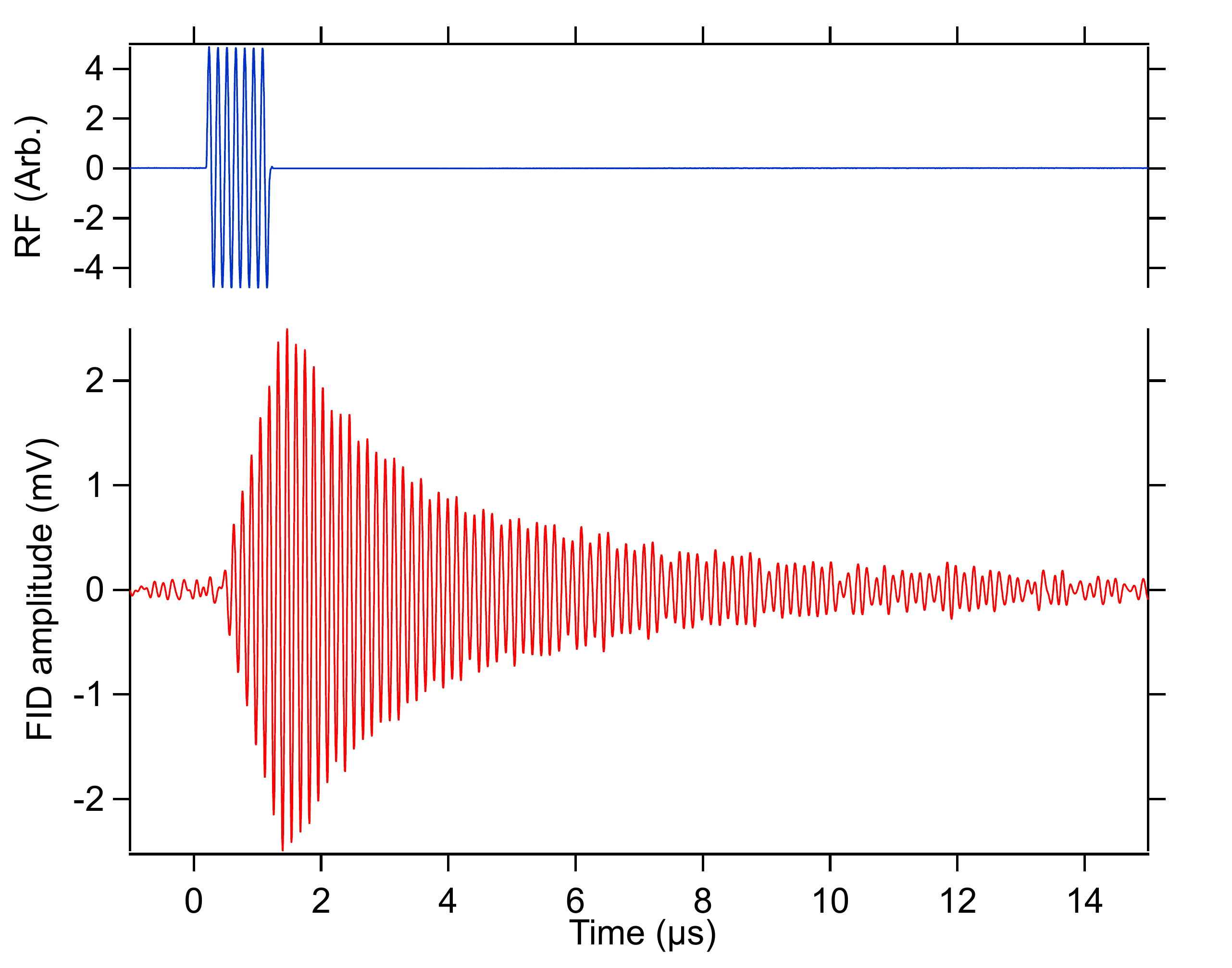}
    \caption{ 
    $^{133}$Cs FID signal, taken at a bias field of 21 G, as described in the text. The FID is excited by a 1~$\mu$s pulse, as shown in the figure. A fit to an exponentially-decaying sinusoid gives a 3~$\mu$s T$_2^*$.
    \label{fig:Cs_FID}}
    \end{center}
\end{figure}

For the case of Cs, we use a mostly RHC pump/probe beam at 846 nm whose intensity and waist are about $10^{3}$ mW/cm$^{2}$ and 200 $\mu$m respectively. This beam passes through the center of the crystal just above the RF (radio-frequency) coil and is subsequently focused onto a fast photo-diode. The RF coil is about 0.5 cm away from the front surface of the crystal. DC bias magnetic fields ranging from a few Gauss to $\sim$ 80 Gauss are applied at $\sim$ 45 degrees relative to the pump beam direction.
We pump the atoms for $\sim$ 150 ms which creates magnetization along the direction of the DC bias field. Then we apply a short (and hence spectrally broadband) RF pulse, which 
induces Larmor precession. We high pass filter the pump beam signal from the photo-diode to obtain the time-varying free-induction-decay signal as shown in Fig. \ref{fig:Cs_FID}.       
Rubidium and potassium are measured in a similar manner.

In the case of Cs, we have made FID measurements over a range of Larmor frequencies from 0.7 to 8~MHz. All return similar values of T$_2^* \approx 3$~$\mu$s. At these fields, the nonlinear Zeeman effect is sufficiently small that the different Larmor precession superposition states are unresolved \cite{arimondo1977experimental}.

\begin{figure}[ht]
    \begin{center}
    \includegraphics[width=\linewidth]{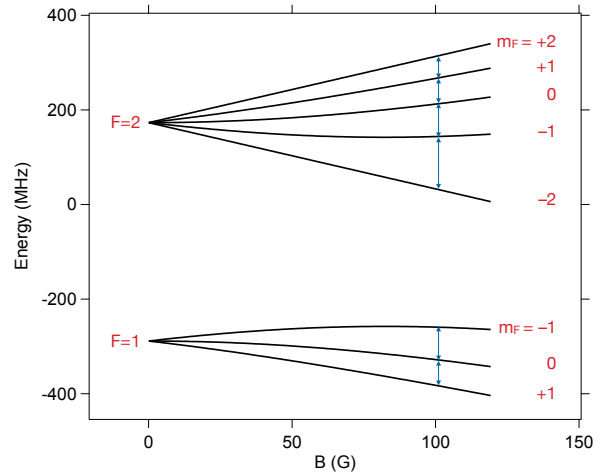}
    \caption{ 
    $^{39}$K Zeeman structure, calculated from reference \cite{tiecke2010properties}. The energy levels are labeled by their low-field quantum numbers. Superposition states of levels differing by $\Delta m_F = 1$ (indicated by arrows) give rise to Larmor precession.
    \label{fig:K39_Zeeman}}
    \end{center}
\end{figure}

\begin{figure}[ht]
    \begin{center}
    \includegraphics[width=\linewidth]{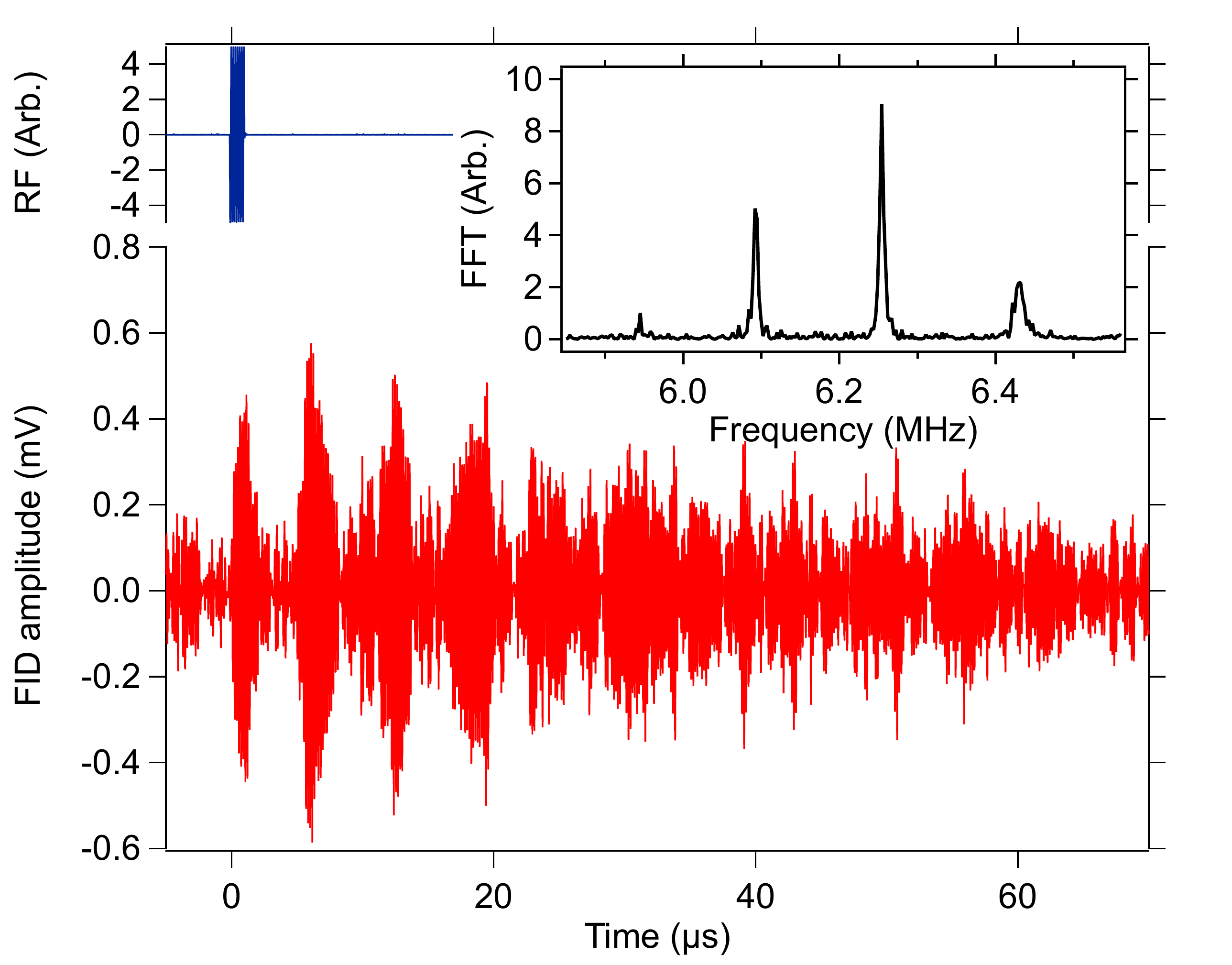}
    \caption{ 
    $^{39}$K FID signal, taken at a bias field of 9 G, as described in the text. On this scale, the individual oscillations of the RF pulse and FID signal are not visible, but their overall envelopes can be observed. The FID signal shows clear beating. The inset shows the Fourier transform (magnitude squared) of the FID signal. Fitting the largest peak to a Lorentzian lineshape gives a 6 kHz full-width-at-half-maximum, which corresponds to a 53~$\mu$s T$_2^*$. From left to right, the four peaks correspond to superpositions of  $|m_F = +2 \rangle $ and $|m_F = +1 \rangle$; $+1$ and $0$; $0$ and $ -1$; and $-1$ and $ -2$. 
%
%
%
    \label{fig:K_FID}}
    \end{center}
\end{figure}

This is not the case for $^{39}$K, whose much smaller hyperfine splitting \cite{arimondo1977experimental} results in a much larger splitting between the different Zeeman states.  The Zeeman structure of $^{39}$K is shown in Fig. \ref{fig:K39_Zeeman}. A typical FID signal for potassium is shown in Fig. \ref{fig:K_FID}. The beating of the different Larmor superposition states makes fitting the decay to a damped sinusoid impractical. Instead, we Fourier transform the FID signal and fit the resulting spectral peaks. From their full width at half-maximum (FWHM), we determine T$_2^*$ from the relationship T$_2^* = ( \pi \cdot \mathrm{FWHM})^{-1}$, where FWHM is expressed in cycles per unit time (e.g. Hz). 
From the spectrum, we determine that the four peaks observed are from the $F=2$ hyperfine manifold of $^{39}$K; the shifts of $^{40}$K, $^{41}$K and the $F=1$ manifold of $^{39}$K are sufficiently large that their Larmor precession transitions would be spectrally resolved \cite{arimondo1977experimental}.

We note that the measured T$_2^*$ for $^{39}$K is over an order of magnitude longer than for $^{133}$Cs. These differences are discussed in section \ref{sec:T2*_vs_species}.

At sufficiently low magnetic fields ($\lesssim 2$  Gauss) Rb FID exhibits a single line, similar to Cs. 
At ``intermediate'' fields, the different Larmor superposition states cannot be fully resolved, but their splitting leads to a decrease in the FID time. At still higher fields ($\gtrsim 40$~Gauss) beating is clearly observed (as in the case of potassium data shown in Fig. \ref{fig:K_FID}).
We present the higher-field data below in section \ref{sec:high_field_Rb_T2*}; for now we concern ourselves with the low-field limit.

We measured the Rb FID time for Rb densities from $10^{17}$ to $10^{18}$~cm$^{-3}$, and saw no variation to within $\pm 15\%$.
%
Similarly, the Rb FID time showed no dependence on the ortho fraction in the crystal over a range from $5\times 10^{-5}$ to $1 \times 10^{-3}$, to within $\pm 10\%$.

Much like Rb, we did not see any dependence of Cs FID decay on Cs density or ortho fraction. We observe no dependence on the Cs density (to within $\pm 15\%$) over the range from $1 \times 10^{16}$ to $1 \times 10^{17}$~cm$^{-3}$. 
We observe no dependence on the ortho fraction (to within $\pm10\%$) over a range from $3 \times 10^{-5}$ to $1 \times 10^{-3}$.
%
For Cs,  T$_2^*$ showed little dependence on the substrate temperature at the time of crystal growth.

We note that for all species, the FID frequency is consistent with the applied magnetic field and the free-atom g-factor \cite{arimondo1977experimental}. However, because we do not know the applied magnetic field  accurately, all we can say is that the g-factor in the crystal matches the free-atom case to within $\pm 20$\%.

\subsection{T$_2^*$ for different species}
\label{sec:T2*_vs_species}

\begin{figure}[ht]
    \begin{center}
    \includegraphics[width=\linewidth]{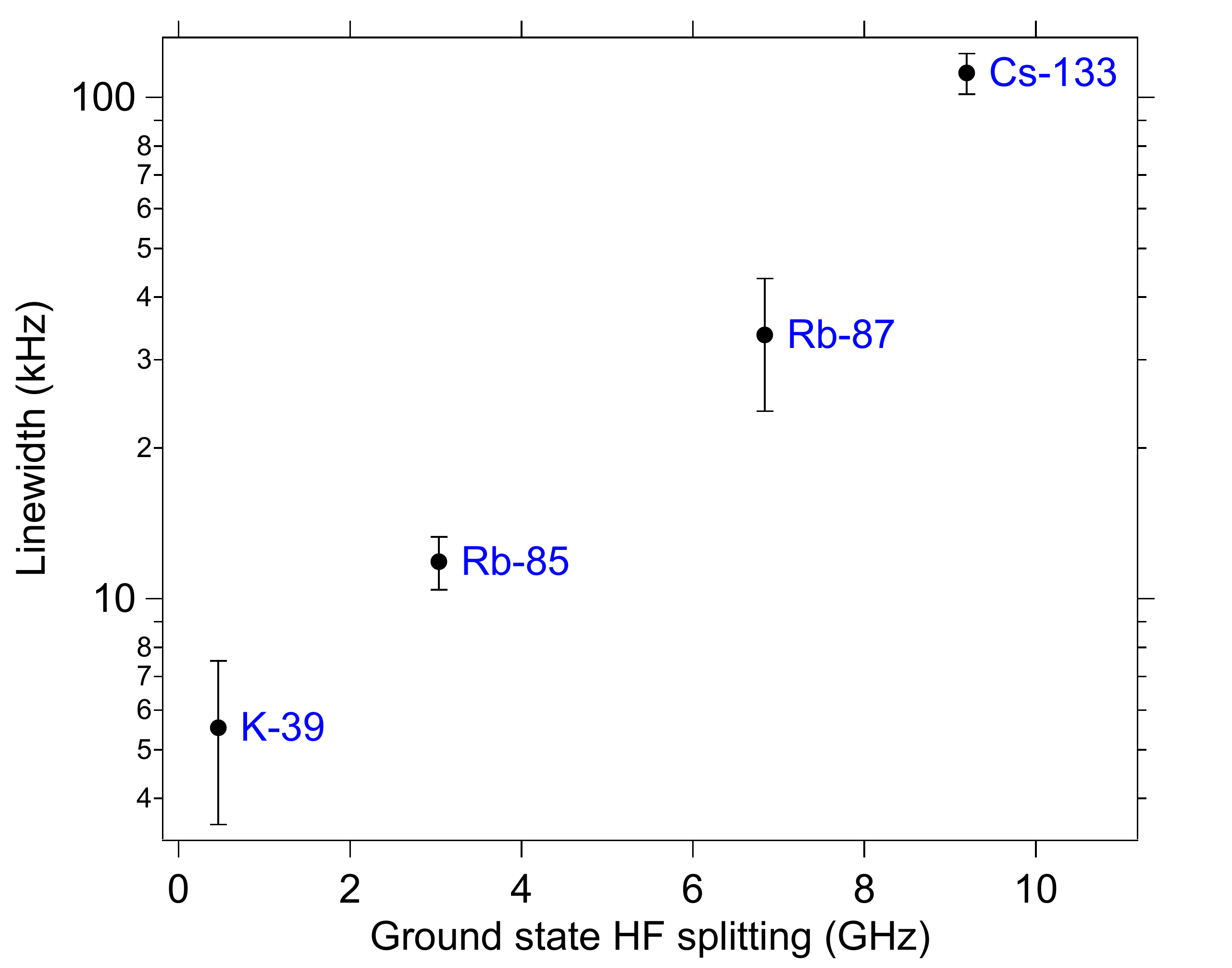}
    \caption{ 
T$_2^*$, expressed as a linewidth $= \frac{1}{\pi \mathrm{T}_2^*}$, for the species measured. The linewidths are plotted as  a function of the ground-state hyperfine splitting of each species; we believe this is the key parameter in explaining the differences in the observed linewidths, as discussed in the text. Rb and Cs data were taken at sufficiently low fields that the different Larmor precession frequencies were unresolved; the $^{39}$K data was taken at similar fields but with resolved structure; the number plotted is the linewidth of the $F=2$, $|m_F = 0\rangle$ and $|m_F=-1\rangle$ Larmor superposition. 
All data points were taken in the short-pulse limit. The unresolved splittings of the different Larmor precession states may be artificially broadening the $^{85}$Rb linewidth at the level of $\lesssim 10\%$; less for $^{87}$Rb amd $^{133}$Cs.
    \label{fig:FID_species_comparison}}
    \end{center}
\end{figure}

Figure \ref{fig:FID_species_comparison} shows the measured T$_2^*$ values, expressed as a FWHM linewidth for our measured species. 

The T$_2^*$ times are limited by inhomogenous broadening, as we have measured spin-echo T$_2$ times to be $\gtrsim 1$~ms for rubidium and cesium (we have not measured spin-echo signals in potassium due to its small polarization signal).

We expect that the inhomogenous broadening which limits T$_2^*$ is primarily due  to electrostatic-like interactions with the host matrix \cite{PhysRevB.100.024106}. As such, we would expect the energy level shifts  to resemble those of the Stark effect. Considering the Stark effect for a ground-state alkali atom, there is a scalar component which shifts all $|F, m_F\rangle$ levels the same, and a tensor component which shifts different $F$ and $m_F$ levels differently. It is this tensor component which will cause inhomogneous broadening for Larmor precession.  The tensor component is zero in second-order perturbation theory, and only appears in third-order perturbation theory including two electric dipole couplings and one hyperfine interaction
\cite{angel1968hyperfine, dzuba2010hyperfine, robyr2014measurement}.
Consequently, we would expect atoms with larger hyperfine splittings to have larger shifts due to their interaction with the matrix. In the case of a polycrystalline matrix with inhomogenous trapping sites, this would result in larger inhomogenous broadening. This is qualitatively consistent with the observations presented in Fig.~\ref{fig:FID_species_comparison}.

A more sophisticated and quantitative model based on the rigorous ESR Hamiltonian  is presented in Section~\ref{subsec:TheoryHyperfine}.

\subsection{T$_2^*$ for different Larmor superposition states}
\label{sec:high_field_Rb_T2*}

\begin{figure}[ht]
    \begin{center}
    \includegraphics[width=\linewidth]{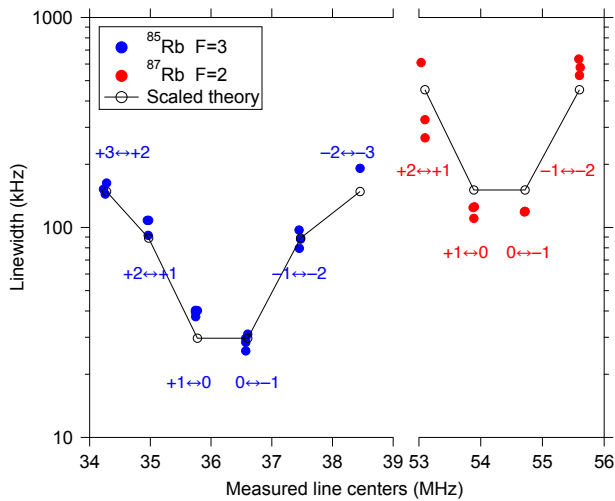} 
    \caption{
    The FWHM linewidths of the Rb Larmor precession superposition states at a magnetic field of 80~G, as discussed in the text. The data are labeled by the $m_F$ states of their corresponding Larmor superposition. Plotted alongside the data is the theory of section \ref{subsec:TheoryHyperfine}, scaled by a factor of 1.35.
    \label{fig:FID_T2*_vs_m}}
    \end{center}
\end{figure}

At sufficiently high magnetic fields, we can spectrally resolve the different Larmor precession states of rubidium, similar to the case of potassium shown in figure \ref{fig:K_FID}.  Figure \ref{fig:FID_T2*_vs_m} shows data for both the $F=3$ manifold of $^{85}$Rb and the $F=2$ manifold of $^{87}$Rb. Larmor precession arises from all superpositions of states that differ by $\Delta m=1$.

As observed in section \ref{sec:T2*_vs_species} at low fields, the $^{87}$Rb linewidths are larger than those of the corresponding superpositions in $^{85}$Rb. 
For both isotopes, the linewidths are larger for superposition states of higher $m_F$. 
Qualitatively, this is as one would expect for inhomogenous broadening from electrostatic interactions: tensor Stark shifts scale as $m_F^2$ \cite{Ulzega:06, dzuba2010hyperfine}.

The data in figure \ref{fig:FID_T2*_vs_m} is presented alongside the quantitative theory of section \ref{subsec:TheoryHyperfine}. The theory reproduces the dependence of the linewidth on both isotope and $m_F$. The  significant isotope effect is mainly due  to the hyperfine anisotropy of $^{87}$Rb, which is 3.4 times larger than that of $^{85}$Rb (see Table \ref{tab:Abinitio}) owing to  the difference in the nuclear magnetic moments. The $m_F$ scaling arises from the tensor nature of the anisotropic hyperfine interaction, as presented in section \ref{subsec:TheoryHyperfine}.

More subtle features of the spectrum, such as why the Larmor precession linewidth of the  $(-1, 0)$ superposition of $F=3$~$^{85}$Rb is consistently narrower than the $(+1,0) $ superposition, are not understood.
The low signal-to-noise of the potassium polarization signal does not permit similar comparisons of different Larmor precession states, and we did not take Cs data at sufficiently high field to resolve the different superpositions.

\subsection{Temperature dependence}

We measured rubidium T$_2^*$ in the low-field limit at different crystal temperatures.
The Rb linewidth showed no dependence on the crystal temperature 
over a range from 3 to 4.2~K, to within $\pm30$\%. 

\begin{figure}[ht]
    \begin{center}
    \includegraphics[width=\linewidth]{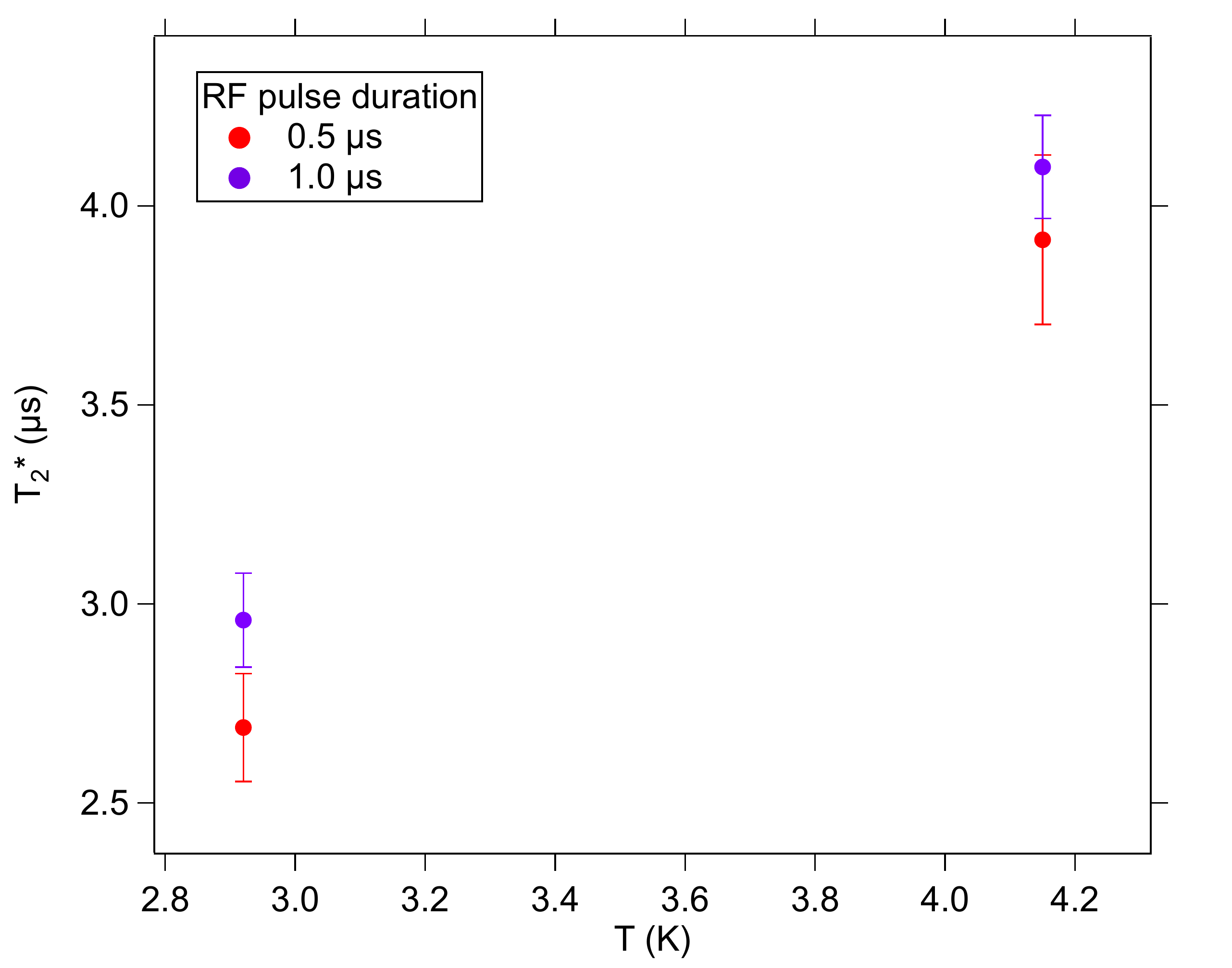}
    \caption{ 
Cs FID T$_2^*$ measured at different temperatures. Each point is an average of multiple measurements over multiple temperature cycles of the same sample; the warmer temperatures consistently gave longer FID decay times. 
    \label{fig:CsFIDvsT}}
    \end{center}
\end{figure}

We do, however, see a dependence of the FID decay time on temperature for Cs.  We warmed a Cs-doped crystal (grown at 3.2~K substrate temperature, with our ``base'' ortho fraction) crystal to 4~K and held it there overnight to allow the crystal to anneal. This produced, surprisingly, longer free-induction decay times by roughly 40\%. Cooling back to our base temperature of 3~K returned our original FID times. Subsequent cycling between 3 and 4~K consistently showed longer FID decay times at the elevated temperature. This data is presented in figure \ref{fig:CsFIDvsT}.

The link between elevated matrix temperature and longer T$_2^*$ times is not understood, but we speculate it may be due to larger amplitude atomic motion (on a timescale much shorter than T$_2^*$)  reducing the anisotropy of individual trapping sites and/or reducing the inhomogeneities between different trapping sites, similar to ``motional narrowing'' effects observed in NMR \cite{PhysRev.73.679}.

\section{Theory}

\subsection{Inhomogeneous broadening due to hyperfine interactions}
\label{subsec:TheoryHyperfine}


In this section, we present a theoretical analysis of inhomogeneous broadening of hyperfine transitions of $^2$S  atoms  embedded in an inert matrix. The theory is  based on the hyperfine  Hamiltonian commonly used to calculate powder ESR spectra \cite{ESR1,Carrington67,slichter1990}, which we extend  to the low-field limit of  interest to the matrix isolation experiments described here. 
The primary focus will be on alkali-metal atoms  trapped in  solid $p$-H$_2$, although our theory is sufficiently general to be applicable to any S-state  atom in an inert matrix. 

To model the broadening of the hyperfine transitions $F,m_F\leftrightarrow F',m_F'$, we assume that  it is due to the tensor matrix shifts of the hyperfine levels caused by the interaction  with the host matrix. As shown below, the tensor matrix shifts depend on the orientation of the principal axes of the hyperfine tensor $\mathbf{A}$ with respect to the magnetic field axis.  We derive analytical expressions for these shifts as a function of the orientation angle and then calculate them for all possible orientations to obtain the linewidth of the hyperfine transitions of an atom  in a polycrystalline (powder) matrix. Our results establish a direct connection between the experimentally observable transition linewidths and the elements of the hyperfine tensor, calculated {\it ab initio} for a range of alkali-H$_2$ complexes as described in section \ref{subsec:Abinitio}. At the end of this section, we compare our calculated transition linewidths with experiment, finding good semi-quantitative agreement, and discuss the limitations of our model.

\begin{figure}[t!]
    \begin{center}
    \includegraphics[width=1.0\linewidth]{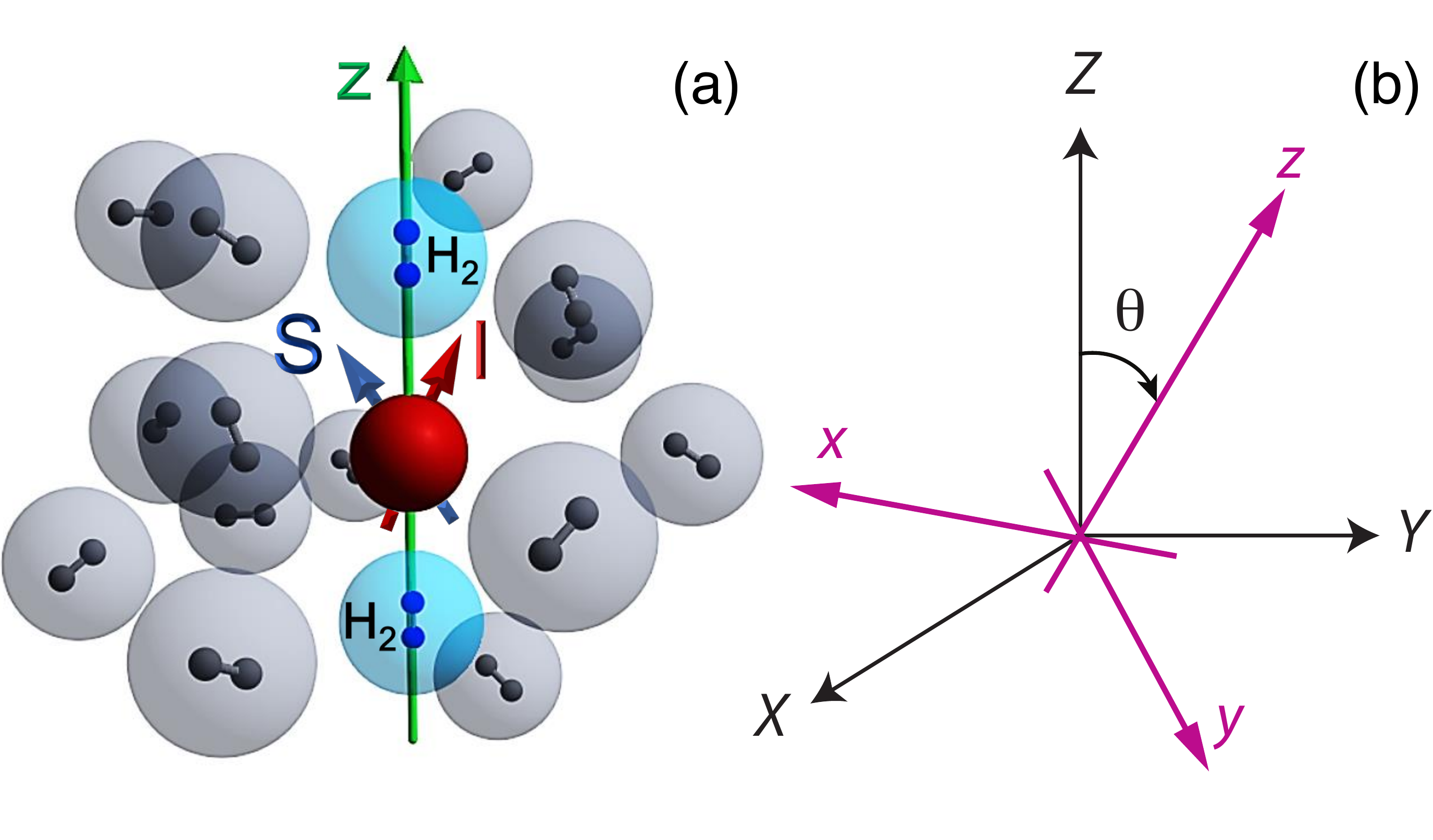} 
    \caption {  (a) A schematic representation of our model for the alkali-metal trapping site in a $p$-H$_2$ matrix. The red circle represents the central alkali-metal atom; the blue circles represent the axial $p$-H$_2$  molecules taken into account in the present calculations; the grey circles are all other $p$-H$_2$ molecules.  The electron and nuclear spins of the alkali-metal atom are indicated by arrows.     (b) Space-fixed (black) and principal-axes (magenta) coordinate systems. The $Z$ axis of the space-fixed system is defined by the direction of the external magnetic field. The positions of the principal axes $x,y,z$ in the space-fixed coordinate system are defined by the Euler angles $\Omega=\phi,\theta,\chi$.   \label{fig:PAsystem} }
    \end{center}
\end{figure}

We begin with the ESR Hamiltonian for a central $S=1/2$ atom embedded in a solid $p$-H$_2$ host matrix  \cite{ESR1,Carrington67,Lund:11}, as illustrated in Fig. \ref{fig:PAsystem}(a)
\begin{equation}\label{Hhf}
H_\text{hf} =   A_{a} \bm{S}\cdot \bm{I} + 2\mu_0  \bm{S}\cdot \mathbf{g} \cdot \bm{B}  + \bm{S}\cdot \mathbf{A} \cdot \bm{I} + \sum_{\alpha} \bm{S} \cdot \mathbf{A}^\alpha \cdot \bm{I}^\alpha,
\end{equation}
where $ \bm{S}$ and $\bm{I} $ are the electron and nuclear spins of the central atom,  $\mathbf{A}$  is the hyperfine  tensor on the central nucleus  of interest, and $\mathbf{A}^\alpha$ are the hyperfine tensors on the surrounding nuclei  bearing nuclear spin angular momenta $\bm{I}^\alpha$ (we neglect this final term in the following calculations). In Eq. (\ref{Hhf}),  $\mathbf{g}$ is the $g$-tensor of the central atom \cite{ESR1,Carrington67,Lund:11},  assumed here to be proportional to the unit matrix,  $\mathbf{g}=g_e\mathbf{1}$, where $g_e\simeq 2$ is the electron $g$-factor.
In defining the hyperfine  tensor, we separate out the contribution due to the hyperfine structure of the free atom $A_{a} \bm{S}\cdot \bm{I}$, which allows us to define unperturbed atomic states  $|F \ m_F\rangle$ in the weak-field limit. Here,   $\bm{F}=\bm{I}+\bm{S}$ is the total angular momentum of the atom, and $m_F$ is the projection of $\bm{F}$ on the space-fixed quantization axis defined by the external magnetic field. 

 The hyperfine tensor  accounts for the modification of the atomic hyperfine structure due to the interaction with the matrix, and can be decomposed as
\begin{equation}\label{Adecomposition}
\mathbf{A} = A_\text{iso}(R)\mathbf{1} + \mathbf{T}
\end{equation}
where the scalar constant $A_\text{iso}$ describes the isotropic (Fermi contact) interaction and the traceless tensor $\mathbf{T}$ describes the anisotropic hyperfine interaction.
Note that the isotropic hyperfine interaction does not affect the splitting between the $m_F$ sublevels of the same $F$-state, so we do not consider this term in the following.
However, it must be taken into account when considering the transitions involving hyperfine states of different $F$.
We further assume that matrix perturbations are weak, {\it i.e.}, $A_{a}\gg  {T}_{ij}$.


The third term in Eq. (\ref{Hhf}) can be written as a sum over Cartesian components of vector operators $ \bm{S}$ and  $ \bm{I} $
\begin{equation}\label{Hhf2}
H_\text{ahf} =   \sum_{i,j=x,y,z}{S}_i T_{ij} I_j 
\end{equation}
In general, the form of this operator depends on the choice of the coordinate system. Here, we choose the principal axes (PA) of the tensor  $\mathbf{T}$ as coordinate axes. The orientation of the PAs with respect to  space-fixed  axes defined by the external magnetic field is specified by the Euler angles $\Omega=(\phi,\theta,\chi)$ as shown in Fig.~\ref{fig:PAsystem}(b).  In this coordinate system, $\mathbf{A}$ and $\mathbf{T}$ take the  diagonal form and Eq. (\ref{Hhf2}) reduces to
\begin{equation}\label{Hhf_PAsys}
H_\text{ahf}^\text{PA} =  T_{xx} S_x I_x +   T_{yy} S_y I_y +  T_{zz} S_z I_z
\end{equation}
where $T_{xx}$, $T_{yy}$, and $T_{zz}$ are the principal axes (PA) components of $\mathbf{T}$ calculated {\it ab initio} as described in the next section.  

In first-order perturbation theory, the energy shift of the atomic level $|Fm_F\rangle$ due to the interaction with the host matrix is given by the diagonal matrix element of the perturbation 
\begin{equation}\label{EnergyShift}
\Delta E_{Fm_F} = \langle Fm_F | H_\text{ahf}^\text{PA}  |Fm_F\rangle 
\end{equation}
To evaluate the matrix elements in Eq. (\ref{EnergyShift})  in terms of the PA components of the hyperfine tensor, we express the Hamiltonian via the spherical tensor operators expressed in the space-fixed frame [see Fig.~\ref{fig:PAsystem}(b)]. Following Appendix A of Ref. \cite{Tscherbul:12} and keeping in mind that  $\bar{T}=\frac{1}{3}(T_{xx}+T_{yy}+T_{zz})=0$, we have
\begin{multline}\label{Hahf_PA_SF}
H_\text{ahf}^\text{PA}  =  \sum_{p=-2}^2 \biggl{[} \frac{1}{2} (T_{xx}-T_{yy})[D^{2}_{p,2}(\Omega) + D^{2}_{p,-2}(\Omega)]  \\  + \frac{1}{\sqrt{6}}(2T_{zz} - T_{xx} - T_{yy})  D^{2}_{p0}(\Omega)   \biggr{]} [\bm{I}\otimes\bm{S}]^{(2)}_p,
\end{multline}
where $[\bm{I}\otimes\bm{S}]^{(2)}_p$ is a rank-2 tensor product of two rank-1 spherical tensor operators and $D^{2}_{p,2}(\Omega)$ are the Wigner $D$-functions of the Euler angles $\Omega$ that define the orientation of  the  PA coordinate system relative to the space-fixed axes [see Fig.~\ref{fig:PAsystem}(b)].

%
In the case of axial symmetry assumed below \cite{Carrington67, Lund:11},  $T_{xx}=T_{yy}$  and the expression (\ref{Hahf_PA_SF}) simplifies to
\begin{equation}\label{EnergyShiftAsialSymm}
H_\text{ahf}^\text{PA}  =  \frac{2}{\sqrt{6}}  \Delta T \sum_{p=-2}^2  D^{2}_{p0}(\Omega)  [\bm{I}\otimes\bm{S}]^{(2)}_p,
\end{equation}
where we define $\Delta T = T_{zz}-T_{xx}$ as the hyperfine anisotropy. The matrix shifts in Eq. (\ref{EnergyShift})  thus become, for a given orientation $\Omega$ of the PA coordinate system relative to the space-fixed axes
\begin{multline}\label{EnergyShift2}
\Delta E_{Fm_F}(\Omega) = \frac{2}{\sqrt{6}}  \Delta T \\ \times \sum_{p=-2}^2   D^{2}_{p0}(\Omega)  \langle Fm_F |  [\bm{I}\otimes\bm{S}]^{(2)}_p  |Fm_F\rangle 
\end{multline}

Applying the Wigner-Eckart theorem \cite{Zare:88} to evaluate the matrix elements on the right-hand side, we find
\begin{multline}\label{matrixElement1}
\langle (IS) Fm_F |  [\bm{I}\otimes\bm{S}]^{(2)}_p  |(IS)F'm_F'\rangle =   (-1)^{F-m_F}  \\ \times
 \threejm{F}{-m_F}{2}{p}{F'}{m_F'} [(2F+1)5(2F'+1)]^{1/2} p_3(I) p_3(S)
 \\  \times  \ninej{I}{I}{1}{S}{S}{1}{F}{F'}{2},
 \end{multline}
 where the symbols in parentheses and curly brackets are 3-$j$ and 9-$j$ symbols, and   $p_3(X) = [(2X+1)X(X+1)]^{1/2}$.
For $m_F=m_F'$,  the 3-$j$ symbol in Eq. (\ref{matrixElement1}) is nonzero only when $q=0$. Setting $D^{2}_{p0}(\Omega)=d^2_{00}(\theta) = \frac{1}{2}(3\cos^2\theta - 1)$ in Eq. (\ref{matrixElement1}), we obtain the angular dependence of the tensor matrix shift 
\begin{multline}\label{matrixElement2}
\Delta E_{Fm_F}(\theta)= \frac{3\cos^2\theta - 1}{\sqrt{6}} \Delta T 
(-1)^{F-m_F} \\ \times
  \threejm{F}{-m_F}{2}{0}{F}{m_F} [(2F+1)5(2F+1)]^{1/2}  \\ \times p_3(I) p_3(S) \ninej{I}{I}{1}{S}{S}{1}{F}{F}{2}
 \end{multline}

For a polycrystalline $p$-H$_2$ matrix, the orientation of the PA coordinate system with respect to the external magnetic field is random \cite{ESR1}, {\it i.e.}, all possible  $\theta$ angles  will contribute to the linewidth. In the presence of axial symmetry, Eq. (\ref{matrixElement1}) shows that there is a distribution of matrix shifts proportional to $(3\cos^2\theta - 1)$.  The transition linewidth can then be evaluated as a difference between the maximum (2) and minimum ($-1$) values of the angular function $(3\cos^2\theta-1)$ for $\theta \in [0,\pi]$. Replacing $3\cos^2\theta-1\to 3$ in Eq. (\ref{matrixElement2}), we obtain the linewidth of the atomic state $|Fm_F\rangle$ in a polycrystalline matrix
\begin{multline}\label{matrixElementFinal}
\Delta E_{Fm_F}=3\sqrt{\frac{5}{6}} \Delta T 
(-1)^{F-m_F} (2F+1)    p_3(I) p_3(S) \\  \times  \threejm{F}{-m_F}{2}{0}{F}{m_F} \ninej{I}{I}{1}{S}{S}{1}{F}{F}{2}.
 \end{multline}
The 3-$j$ symbol on the right is equal to  $ (-1)^{F-m_F}[(2F+3)(2F+2)(2F+1)2F(2F-1)]^{-1/2} [3m_F^2 - F(F+1)]$ \cite{Zare:88}. Equation (\ref{matrixElementFinal}) thus establishes that for a given alkali-metal atom (fixed $I$, $S$, and $\Delta T$), the linewidth of the $F,\,m_F$ level scales with $F$ and $m_F$ as
\begin{multline}\label{Fscaling}
\Delta E_{Fm_F}\propto [(2F+3)(2F+2)(2F+1)2F(2F-1)]^{-1/2} \\ \times (2F+1) \ninej{I}{I}{1}{S}{S}{1}{F}{F}{2} [3m_F^2 - F(F+1)]
 \end{multline}

Given the broadening of the individual hyperfine levels (\ref{matrixElementFinal}), we can calculate the inhomogeneous transition linewidth assuming that the hyperfine levels $F,\, m_F$ and $F',\,m_F'$  involved in the magnetic dipole transition are broadened by the interaction with the matrix. 
Both of the hyperfine levels experience tensor matrix shifts according to Eq. (\ref{matrixElementFinal}). Taking the difference of the $F,m_F$ and $F',m_F'$ level shifts given by Eq.~(\ref{matrixElementFinal}) and averaging the result over $\theta$ as described above, we obtain the  inhomogeneous linewidth of the $F,m_F \leftrightarrow  F',m_F'$ transition 
\begin{multline}\label{transitionEnergyShiftGeneral}
 \Delta E_{Fm_F \leftrightarrow F'm_F'} =3{\sqrt{5/6}}  p_3(I) p_3(S)  \Delta T
  \\   \times
\Biggl{[}   (-1)^{F-m_F} (2F+1)   \threejm{F}{-m_F}{2}{0}{F}{m_F}\ninej{I}{I}{1}{S}{S}{1}{F}{F}{2} \\ 
    - (-1)^{F'-m_F'} (2F'+1)   \threejm{F'}{-m_F'}{2}{0}{F'}{m_F'} \ninej{I}{I}{1}{S}{S}{1}{F'}{F'}{2} \Biggr{]}.
 \end{multline} 
 For the transitions involving different $m_F$ sublevels of the same $F$ state of interest here,  $F=F'$ and Eq.~(\ref{transitionEnergyShiftGeneral})  simplifies to [omitting the irrelevant overall phase $(-1)^{F-m_F}$]
 \begin{multline}\label{transitionEnergyShiftSameF}
\Delta E_{Fm_F \leftrightarrow F'm_F'}=3{\sqrt{5/6}} p_3(I) p_3(S)  (2F+1)  \Delta T  \\ 
\times \Biggl{[}   \threejm{F}{-m_F}{2}{0}{F}{m_F} - (-1)^{m_F-m_F'}  \threejm{F}{-m_F'}{2}{0}{F}{m_F'}  \Biggr{]}  \\
\times   \ninej{I}{I}{1}{S}{S}{1}{F}{F}{2}
 \end{multline}

 \begin{table}
	\caption{Calculated  linewidths (in kHz) for the $F,m_F \leftrightarrow F,m_F'$ transitions in different alkali-metal  isotopes. The theoretical values are computed using Eq. (\ref{transitionEnergyShiftSameF}) based on the {\it ab initio} hyperfine anisotropies $\Delta T$ calculated as described in  Sec.~\ref{subsec:Abinitio}.  The theoretical $m_F\leftrightarrow m_F'$ transition linewidths are invariant with respect to the simultaneous sign reversal $m_F\to -m_F$ and $m_F'\to -m_F'$; thus only positive values are presented.}
	\vspace{0.3cm}
	 \begin{tabular}{cc}
	 \hline
	 	 \hline
	  Transition  ($m_F\leftrightarrow m_F'$) &  Theory    \\ \hline	  
	  	 \multicolumn{2}{l}{$^{39}$K, $F=2$}\\  \hline
	$2\leftrightarrow 1$  &   6.19        \\
	$1\leftrightarrow 0$  &   2.06    \\
	       \hline
	  	 \multicolumn{2}{l}{ $^{85}$Rb, $F=3$} \\  \hline
	$3\leftrightarrow 2$  &   109.6          \\
    	$2\leftrightarrow 1$  &   65.78         \\
    	$1\leftrightarrow 0$  &   21.93         \\	
       	       	 \hline
	  	 \multicolumn{2}{l}{ $^{87}$Rb, $F=2$} \\  \hline	
	       	 \hline
    	$2\leftrightarrow 1$  &  334.4        \\
    	$1\leftrightarrow 0$  &   111.5       \\	
	       \hline
	  	 \multicolumn{2}{l}{ $^{133}$Cs, $F=4$} \\  \hline
	$4\leftrightarrow 3$  &   546.27      	\\ 
	$3\leftrightarrow 2$  &   390.18             \\
    	$2\leftrightarrow 1$  &   234.11             \\
    	$1\leftrightarrow 0$  &   78.04      \\	
       	       	 \hline
	 \end{tabular}
	\label{tab:Theory_vsExp}
\end{table}

Table \ref{tab:Theory_vsExp} presents the theoretical linewidths of $m_F$-changing transitions in different alkali-metal atoms. The linewidths are calculated using Eq. ~(\ref{transitionEnergyShiftSameF}) based on the {\it ab initio} values of the hyperfine anisotropy $\Delta T$ from Sec. \ref{subsec:Abinitio}. We observe  good semi-quantitative agreement between theory and experiment across all species and isotopes,  confirming that anisotropic hyperfine interactions are the dominant source of broadening. 

The overall trend of the measured linewidths to increase from K to Rb and from Rb to Cs is well reproduced by the theory.  The reason for this trend is that the calculated linewidths (\ref{transitionEnergyShiftSameF}) are proportional to the hyperfine anisotropy $\Delta T$, which increases in the sequence K $\to$ Rb $\to$ Cs (see Table \ref{tab:Abinitio}). The small magnitude of the K linewidths is a result of its exceedingly small hyperfine anisotropy, which is a factor of 10 smaller than the values calculated for Rb and Cs complexes.

For the same alkali-metal isotope, Eq. (\ref{transitionEnergyShiftSameF}) predicts $F$-independent broadening of  the $F,m_F\leftrightarrow F,m_F'$ transitions. 
Within the same $F$-manifold, the linewidths are expected to increase linearly with $m_F$ and to be independent of its sign, again consistent with the trend observed experimentally (Fig.~\ref{fig:FID_T2*_vs_m}).  Significantly,  Eq.~(\ref{Fscaling}) predicts that $ + m_F \leftrightarrow -m_F$ transitions will have dramatically reduced inhomogenous broadening, as these pairs of levels are (to first order) shifted identically by 
%
the anisotropic hyperfine interaction. 
Experimentally, such transitions are found to have much smaller linewidths that the Larmor-precession transitions, as discussed in Sec. \ref{sec:TheoryGeneric} \cite{PhysRevB.100.024106}.


While our theoretical results are in nearly quantitative agreement with experiment, small disagreements remain. We suspect these disagreements are due to differences between our model trapping site and the true trapping site. 
To compensate for this, we scale our theoretical Rb anisotropies by a single constant factor (common to both isotopes). This scaled calculation is presented alongside experimental data in figure \ref{fig:FID_T2*_vs_m}. With this scaling, we see nearly quantitative agreement with experiment.

Additional work is warranted to provide more detailed models of trapping sites, which are different not only in their orientations, but also in their geometries and coordination numbers \cite{scharf1993nature}, bringing about additional broadening mechanisms. A theoretical study of these mechanisms would require a detailed investigation of trapping site structure (using, {\it e.g.}, quantum Monte Carlo simulations) combined with extensive {\it ab initio} calculations of the hyperfine and $g$-tensor elements corresponding to different site structures.





\subsection{Ab initio calculations of alkali-H$_2$ potentials and hyperfine interactions}
\label{subsec:Abinitio}

As discussed in Sec. \ref{subsec:TheoryHyperfine}, the linewidths of alkali-metal atoms trapped in solid $p$-H$_2$ are determined by the hyperfine anisotropy $\Delta T$. To estimate this quantity, we adopt a minimal model for the alkali-metal trapping site illustrated in figure \ref{fig:PAsystem}. In this axially symmetric model, commonly used in theoretical simulations of molecular ESR spectra \cite{Carrington67,Lund:11}, the central alkali-metal atom A is surrounded by two H$_2$ molecules in the linear configuration H$_2$--A--H$_2$. We then use the eigenvalues of the hyperfine tensor calculated {\it ab initio} at  the equilibrium A--H$_2$ geometry $R_e$ to approximate the hyperfine anisotropy $\Delta T$ defined in Sec. \ref{subsec:TheoryHyperfine} above. 

\begin{figure}[t]
    \begin{center}
    \includegraphics[width=\linewidth]{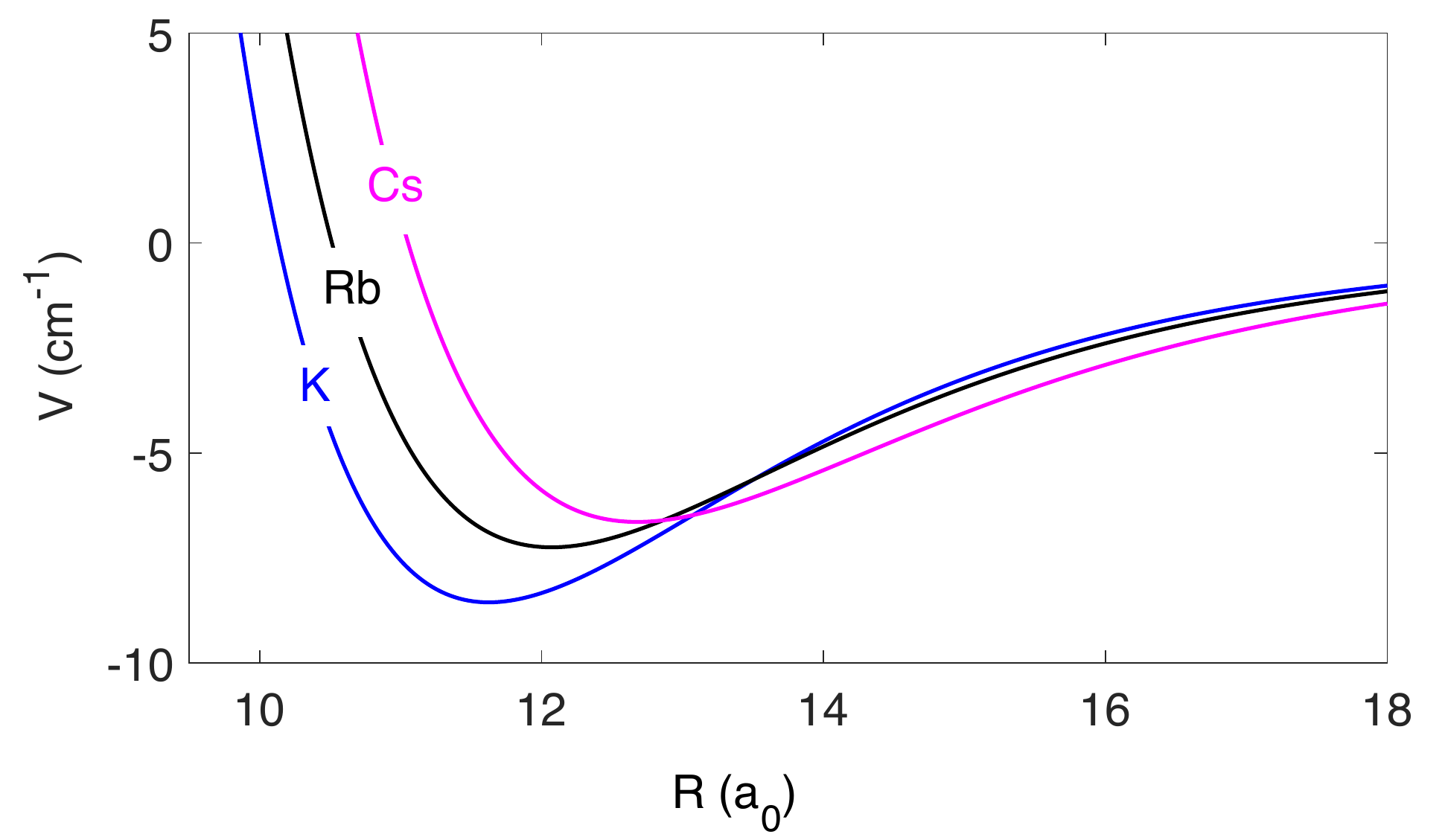}
    \caption{ 
  {\it Ab initio} isotropic interaction potentials for  K, Rb, and Cs atoms with H$_2$.
    \label{fig:TheoryPotentials}}
    \end{center}
\end{figure}

To estimate the equilibrium configuration of the axial trapping site, we carried out {\it ab initio} calculations of the alkali-H$_2$ interaction potentials using the unrestricted coupled cluster method with singles, doubles and perturbative triples  [UCCSD(T)]  \cite{SevaAbinitio1}, as implemented in \mbox{MOLPRO} \cite{SevaAbinitio2}. The aug-cc-pVQZ \cite{SevaAbinitio3} and Jorge-AQZP \cite{SevaAbinitio4} one-electron basis sets were employed for H and K atoms, respectively. 
For Rb and Cs atoms, $n$ core electrons were replaced with the ECPnMDF relativistic effective potential ($n=28$ for Rb and $n=46$ for Cs). The remaining valence 
electrons of Rb and 
Cs were described with the uncontracted $[13s10p5d3f]$ and $[12s11p5d3f]$ basis sets \cite{SevaAbinitio5}, respectively. The alkali-H$_2$ interaction potentials were corrected for the basis set superposition error \cite{SevaAbinitio6} and expressed in Jacobi coordinates $R$ and $\theta$, where $R$ is the interatomic distance between an A atom and the H$_2$ center of mass, and $\theta$ is the angle between the A-H$_2$ vector $R$ and the H$_2$ interatomic axis. The two-dimensional interaction energies were averaged over 19 equally spaced values of $\theta \in [0,90^\circ]$ using the hindered rotor model \cite{li2010adiabatic} and fitted with cubic splines to produce the isotropic potentials shown in Fig.~\ref{fig:TheoryPotentials}.

As shown in Sec.~\ref{subsec:TheoryHyperfine}, the hyperfine tensor on the nucleus of interest has the isotropic ($A_\text{iso}\mathbf{1}$) and anisotropic ($\mathbf{T}$) components, which can be expressed as the Fermi contact and spin-dipolar terms in SI units:
\begin{equation}\label{Aiso}
A_\text{iso}= \frac{g_N e^2 \hbar}{6\pi \epsilon_0 c^2 m_e m_p} |\Psi(\mathbf{r})|^2,
\end{equation}
\begin{equation}
\mathbf{T}= \frac{g_N e^2 \hbar}{16\pi^2  \epsilon_0 c^2 m_e m_p} \biggl{\langle} \frac{\mathbf{r}^t\cdot \mathbf{r} \cdot \mathbf{1} - 3\mathbf{r}  \cdot \mathbf{r}^t }{r^3}\biggr{\rangle},
\end{equation}
where $g_N$ is the nuclear $g$-factor, $e$ is the electron charge, $\hbar$ is the reduced Planck constant, $\epsilon_0$ is vacuum permittivity, $c$ is the speed of light, $m_e$ and $m_p$ are the electron and proton masses, $|\Psi(\mathbf{r})|^2$ is the electron spin density at the nucleus, and the expectation value $\langle ...\rangle$  is that of the spin-dipolar interaction.  
We carried out {\it ab initio} calculations of the spin density $|\Psi(\mathbf{r})|^2$ and the spin-dipolar interaction on the alkali-metal nucleus using the UCCSD(T) method and all-electron fully uncontracted basis sets augmented by the large-exponent $s$ functions in CFOUR \cite{SevaAbinitio8}. The aug-cc-pwCV5Z \cite{SevaAbinitio9} and relativistic ANO-RCC \cite{SevaAbinitio10} basis sets augmented with four $s$ functions obtained by multiplying the largest exponent by a factor of 4 were used for H and alkali-metal atoms, respectively, as described in our previous work on alkali-He hyperfine interactions \cite{Tscherbul:09,Tscherbul:11}. We carried out test calculations of the hyperfine tensor for $^1$H, $^{39}$K, $^{85}$Rb, $^{87}$Rb and $^{133}$Cs  with the corresponding nuclear spins $I = 1/2$, 3/2, 5/2, 3/2 and 7/2.

To validate the level of theory used to predict the anisotropic component of the hyperfine  tensor, we also calculated its isotropic component $A_\text{iso}$ in Eq.~(\ref{Aiso}). Table \ref{tab:AbinitioAiso} compares the calculated and experimental values of the hyperfine constants for 	$^{1}$H and the alkali-metal atoms. 
For the light $^1$H, $^7$Li, and $^{39}$K isotopes, the calculated and experimental values are in good agreement. For Rb isotopes, we observe significant deviations from experiment because of the relativistic properties of the core electrons, which are not accounted for in our {\it ab initio} calculations. It is important to note that the isotropic part of hyperfine interaction depends on the electron density at a nucleus, while the anisotropic part is defined by the spin-dipolar interaction, which is much less affected by the electron density of the core electrons. Therefore, we expect a much higher accuracy in our anisotropic hyperfine constant calculations on heavy alkali-metal isotopes.  

 \begin{table}
	\caption{Calculated isotropic hyperfine interaction constants (in MHz) compared with experiment for atomic hydrogen (Ref.~\cite{SevaAbinitio11}) and alkali-metal atoms (Ref.~\cite{arimondo1977experimental}). }
	\vspace{0.3cm}
	 \begin{tabular}{ccc}
	 \hline
	 	 \hline
	  Atom  &  This work &   Experiment   \\  \hline	  
	$^{1}$H  &   1418  &   1420.405 726(3)    \\
	$^{7}$Li	&   399	&     401.752 043 3(5)  \\
	$^{39}$K   &	221	  &   230.859 860 1(3)  \\
	$^{85}$Rb    & 848  &	1011.910 813(2)   \\
	$^{87}$Rb    & 2875  &      3417.341 306 42(15)    \\	
       	       	 \hline
	 \end{tabular}
	\label{tab:AbinitioAiso}
\end{table}

 \begin{table}
	\caption{Principal-axis components $(T_{xx},T_{yy}, T_{zz})$ of the hyperfine tensor (in kHz) for the H$_2$--A--H$_2$ complexes. The hyperfine anisotropy $\Delta T = T_{zz}-T_{xx}$. The value of $R$ is fixed at the equilibrium distance $R_e$ of the corresponding  A-H$_2$ interaction potential (see Fig.~\ref{fig:TheoryPotentials}). }
	\vspace{0.3cm}
	 \begin{tabular}{cc}
	 \hline
	 	 \hline
	  System  &   $(T_{xx},T_{yy}, T_{zz})$    \\  \hline	  
	H$_2$--$^{39}$K--H$_2$  &   (-1.8, -1.8, 3.7)   \\
	H$_2$--$^{85}$Rb--H$_2$  &  (-29.2, -29.2, 58.5)   \\
	H$_2$--$^{87}$Rb--H$_2$  &  (-99.1, -99.1, 198.2)   \\	
	H$_2$--$^{133}$Cs--H$_2$  &  (-138.7, -138.7, 277.5)   \\	
       	       	 \hline
	 \end{tabular}
	\label{tab:Abinitio}
\end{table}

 Figure~\ref{fig:TheoryPotentials} shows the radial dependence of the isotropic part of our {\it ab initio} alkali-H$_2$ interaction potentials.
We note that the potential minima of all alkali-H$_2$ complexes occur at much larger distances than the H$_2$--H$_2$ potential minimum, and also they are much larger than the 7~$a_0$ nearest-neighbor spacing in  zero-pressure solid hydrogen \cite{RevModPhys.52.393}. This ``mismatch'' in sizes may explain the existence of multiple trapping sites in the solid \cite{scharf:9013}, as there may be multiple different configurations of similar (or lower) energy than a simple interstitial or  single-substitution site.
The well depths of the potentials are $D_e = -8.5$~cm$^{-1}$ at $11.7~a_0$ for K--H$_2$, $D_e = -7.2$~cm$^{-1}$ at 12.1~$a_0$ for Rb--H$_2$, and $D_e = -6.6$~cm$^{-1}$ at 12.5~$a_0$ for Cs--H$_2$. 

 In Table~\ref{tab:Abinitio}, we report the values of anisotropic components of the hyperfine tensor for the  linear H$_2$--A--H$_2$ complex at the equilibrium A--H$_2$ separation determined from the {\it ab initio} potentials plotted in Fig.~\ref{fig:TheoryPotentials}. In these calculations, the H$_2$ bond is taken to be collinear 
 to the symmetry axis of the axially symmetic H$_2$--A--H$_2$  complex.
We estimate the upper limits to the  hyperfine anisotropy $\Delta T= T_{zz}-T_{xx}$ to be 5.5, 87.7, 297.3 and 416.2 kHz for $^{39}$K, $^{85}$Rb, $^{87}$Rb, and $^{133}$Cs, respectively. 


\section{Properties of inhomogeneous broadening from generic time-symmetric perturbations}
\label{sec:TheoryGeneric}

Our measured T$_2^*$ times for Larmor precession states agree well with the theoretical model for inhomogenous broadening due to hyperfine interactions with an inhomogenous host matrix, as presented in section \ref{subsec:TheoryHyperfine}. First-order perturbation theory --- in the limit that $F$ and $m_F$ are good quantum numbers --- finds that  states of the same $F$ and $|m_F|$ undergo identical shifts. This will be the case not only for the specific interaction Hamiltonian used in section \ref{subsec:TheoryHyperfine}, but for any electrostatic-like perturbation (i.e. a perturbation which is unchanged under time reversal).
%

Because electrostatic interactions are unchanged under time reversal, the electrostatic shift of the $|F,m_F\rangle$ and the $|F,-m_F\rangle$ level should be the same to first order. Hence, superpositions of such levels should show dramatically reduced broadening when compared to Larmor precession levels. This effect has been demonstrated in previous measurements of $^{85}$Rb in parahydrogen \cite{PhysRevB.100.024106}. 
We wish to consider the specific behavior of this phenomena in greater detail here, and compare the broadening of different superposition states.

We first construct a Hamiltonian for the known gas-phase hyperfine and Zeeman structure of the ground state of $^{85}$Rb ($I=5/2$), working in the 12-dimensional subspace of the $^2S_{1/2}$ electronic ground state \cite{steck2013rubidium85}.
We  model the crystal field interaction as a random Hermitian matrix in this subspace, with each element a Gaussian distribution of amplitudes chosen to roughly match our observed T$_2^*$. We then ``time-symmetrize'' the matrix by adding it to a time-reversed copy of itself.
We solve for the eigenvalues of the total Hamiltonian, calculate the energy differences between each pair of levels (labelled by their low-field, perturbation-free eigenvalues), and then repeat the process multiple times and calculate the standard deviation of the distribution of energy differences.

\begin{figure}[ht]
    \begin{center}
    \includegraphics[width=\linewidth]{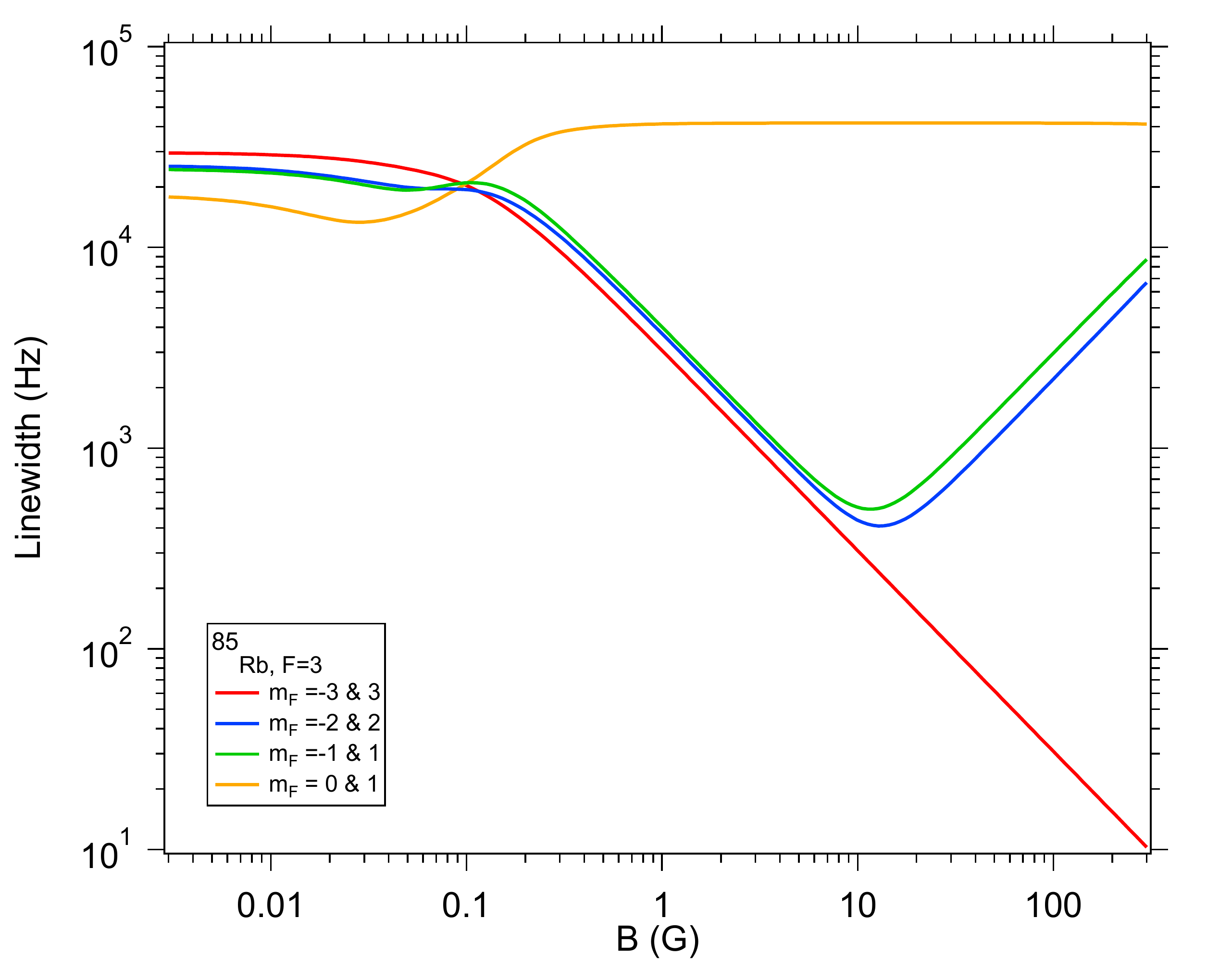}
    \caption{ 
Simulation of inhomogenous broadening for different superpositions of Zeeman levels of the $F=3$ manifold of $^{85}$Rb, plotted as a function of magnetic field. The simulations were for ``random matrix'' perturbations that were symmetric under time-reversal (i.e. electrostatic-like), as explained in the text. The states are labelled by $F$ and $m_F$ quantum numbers; we note that these are good quantum numbers only in the limit of small magnetic fields (Zeeman shifts small compared to the hyperfine splitting) and small random matrix perturbations (small compared to the Zeeman splitting).
    \label{fig:Rb_T2*_sim}}
    \end{center}
\end{figure}

This simple model will capture some of the generic effects of a time-symmetric perturbation, but will miss many of the important elements of our inhomogenous broadening. The model omits the specific structure and symmetry of the trapping sites. It also emits the specific nature of the electrostatic interactions (which will cause different shifts for different $m_F$ levels and different species, as discussed in sections \ref{sec:T2*_vs_species}, \ref{sec:high_field_Rb_T2*}, and \ref{subsec:TheoryHyperfine}). Additionally, it has no predictive capability for the magnitude of the broadening, as the magnitude of the random matrix elements are chosen to match experiment. However, it does reveal interesting behavior which we expect will be general, as shown in Fig. \ref{fig:Rb_T2*_sim}, which plots the simulated linewidths as a function of the applied magnetic field, for Zeeman shifts small compared to the hyperfine splitting.

As shown in Fig. \ref{fig:FID_T2*_vs_m}, $^{85}$Rb ($F=3$) has six superpositions which give rise to Larmor precession. All show roughly similar behavior in this simple calculation; in Fig. \ref{fig:Rb_T2*_sim} we have plotted the linewidth of a single superposition to simplify the graph. All show a linewidth which is roughly independent of the applied field. More interesting is the behavior of the other states shown. The $m_F=+3$ and $-3$ levels are, in the absence of the crystal field, time reversals of each other. At high magnetic fields, where the Zeeman splitting is much greater than the crystal field interaction, this leads to a large suppression of the inhomogenous broadening, as the perturbation by the crystal field is zero to first order. At lower fields, this ``protection'' is reduced as higher-order perturbations begin to play a larger role. In the low-field limit, where the crystal field is greater than the Zeeman splitting, the levels are fully mixed by the crystal field perturbation and the protection is lost, as seen in Fig. \ref{fig:Rb_T2*_sim}.
Slightly more complex are the $m_F$, $-m_F$ superpositions which are not stretched states ($+2$ and $-2$, and $+1$ and $-1$ for $^{85}$Rb). In the absence of the crystal field, these states are time-reversals of each other only in the low-magnetic-field limit. Hence, they show behavior similar to the streched-state superposition at low magnetic fields, but at higher magnetic fields lose their ``protection''  due to the nonlinear Zeeman effect.

These simulations qualitatively agree with our experimental measurements. First, we typically find that our optical pumping signal is reduced in amplitude if we work at longitudinal magnetic fields $\lesssim 1$~Gauss, as discussed in section \ref{sec:B_field_dependence_of_P}. This is qualitatively consistent with the idea that the $m_F$ levels are fully mixed by the crystal-field perturbation at low magnetic fields.
Second, in reference \cite{PhysRevB.100.024106}, we measured the linewidth of a superposition of $|F=3, m_F=+1\rangle$ and $|F=3, m_F=-1\rangle$ at magnetic fields ranging from 60 to 150~G. The linewidth observed was significantly narrower than any of the Larmor precession superpositions. The linewidth increased linearly with the magnetic field over the measured range, in qualitative agreement with the model shown in figure \ref{fig:Rb_T2*_sim}. From the simulation, we expect  significant improvements could be obtained by working with a superposition of stretched states, and at higher fields.

\section{Discussion}

The optical spin polarization signals obtained for Rb and Cs in solid parahydrogen are significantly larger than had previously been reported for alkali atoms in solid argon or neon \cite{kanagin2013optical, pawanthesis}, but not as large as what has been observed in solid helium \cite{PhysRevA.60.3867}. However, the behavior observed in section \ref{sec:PolVsGrowth}  suggests that significant improvement could be obtained in an apparatus capable of colder substrate temperatures and higher parahydrogen deposition rates.

The measured T$_2^*$ times are significantly shorter to those of cesium atoms in solid helium \cite{Weis1996}, but are predominantly due to matrix inhomogeneities. Significant improvement would be observed with a sample of uniform trapping sites in a single-crystal hydrogen matrix. Even in the absence of uniform trapping sites,  significant gains in the spin T$_2^*$ would be expected  by employing stretch-state superpositions rather than Larmor precession states \cite{PhysRevB.100.024106}, as discussed in Section \ref{sec:TheoryGeneric}.

Considering the other alkali metal atoms, we would expect lithium and sodium to have long T$_2^*$ times due to their small hyperfine splitting \cite{arimondo1977experimental}, as explained in sections \ref{sec:T2*_vs_species} and \ref{subsec:TheoryHyperfine}. Unfortunately, we would expect poor polarization signals from lithium and sodium due to their small excited state fine structure splitting, as discussed in section \ref{subsec:OPspecies}.

Considering other elements of the periodic table, we expect that --- unless one is able to grow single-crystal samples with uniform trapping sites --- atoms with ground states with $J > 1/2$ will have short T$_2^*$ times, as tensor Stark shifts would be expected to be significantly larger. Among the $J=1/2$ elements, silver appears promising: it has a large excited-state fine structure splitting (which should be favorable for optical pumping and readout of spin) and a small ground-state hyperfine splitting (which should be favorable for a long T$_2^*$). Moreover, silver's nuclear spin of $I=1/2$ makes it straightforward to obtain stretched-state superpositions with a simple two-photon transition \cite{PhysRevB.100.024106}. In addition, silver's smaller ``size'' may allow it to fit into the lattice in a more stable or favorable configuration.

\section*{Acknowledgements}
This material is based upon work supported by the National Science Foundation under Grants No. PHY-1607072, PHY-1912425, PHY-1607610,  PHY-1912668, and CHE-1654547.
We gratefully acknowledge helpful conversations with Andrei Derevianko, Amar Vutha,  Pierre-Nicholas Roy, and Peng Zhang.
\bibliography{pH2_alkali_comparison2019.bib}

\begin{thebibliography}{68}%
\makeatletter
\providecommand \@ifxundefined [1]{%
 \@ifx{#1\undefined}
}%
\providecommand \@ifnum [1]{%
 \ifnum #1\expandafter \@firstoftwo
 \else \expandafter \@secondoftwo
 \fi
}%
\providecommand \@ifx [1]{%
 \ifx #1\expandafter \@firstoftwo
 \else \expandafter \@secondoftwo
 \fi
}%
\providecommand \natexlab [1]{#1}%
\providecommand \enquote  [1]{``#1''}%
\providecommand \bibnamefont  [1]{#1}%
\providecommand \bibfnamefont [1]{#1}%
\providecommand \citenamefont [1]{#1}%
\providecommand \href@noop [0]{\@secondoftwo}%
\providecommand \href [0]{\begingroup \@sanitize@url \@href}%
\providecommand \@href[1]{\@@startlink{#1}\@@href}%
\providecommand \@@href[1]{\endgroup#1\@@endlink}%
\providecommand \@sanitize@url [0]{\catcode `\\12\catcode `\$12\catcode
  `\&12\catcode `\#12\catcode `\^12\catcode `\_12\catcode `\%12\relax}%
\providecommand \@@startlink[1]{}%
\providecommand \@@endlink[0]{}%
\providecommand \url  [0]{\begingroup\@sanitize@url \@url }%
\providecommand \@url [1]{\endgroup\@href {#1}{\urlprefix }}%
\providecommand \urlprefix  [0]{URL }%
\providecommand \Eprint [0]{\href }%
\providecommand \doibase [0]{http://dx.doi.org/}%
\providecommand \selectlanguage [0]{\@gobble}%
\providecommand \bibinfo  [0]{\@secondoftwo}%
\providecommand \bibfield  [0]{\@secondoftwo}%
\providecommand \translation [1]{[#1]}%
\providecommand \BibitemOpen [0]{}%
\providecommand \bibitemStop [0]{}%
\providecommand \bibitemNoStop [0]{.\EOS\space}%
\providecommand \EOS [0]{\spacefactor3000\relax}%
\providecommand \BibitemShut  [1]{\csname bibitem#1\endcsname}%
\let\auto@bib@innerbib\@empty
\bibitem [{\citenamefont {B{\"u}ch}\ \emph {et~al.}(2013)\citenamefont
  {B{\"u}ch}, \citenamefont {Mahapatra}, \citenamefont {Rahman}, \citenamefont
  {Morello},\ and\ \citenamefont {Simmons}}]{buch2013spin}%
  \BibitemOpen
  \bibfield  {author} {\bibinfo {author} {\bibfnamefont {H.}~\bibnamefont
  {B{\"u}ch}}, \bibinfo {author} {\bibfnamefont {S.}~\bibnamefont {Mahapatra}},
  \bibinfo {author} {\bibfnamefont {R.}~\bibnamefont {Rahman}}, \bibinfo
  {author} {\bibfnamefont {A.}~\bibnamefont {Morello}}, \ and\ \bibinfo
  {author} {\bibfnamefont {M.}~\bibnamefont {Simmons}},\ }\href@noop {}
  {\bibfield  {journal} {\bibinfo  {journal} {Nature communications}\ }\textbf
  {\bibinfo {volume} {4}} (\bibinfo {year} {2013})}\BibitemShut {NoStop}%
\bibitem [{\citenamefont {Tyryshkin}\ \emph {et~al.}(2012)\citenamefont
  {Tyryshkin}, \citenamefont {Tojo}, \citenamefont {Morton}, \citenamefont
  {Riemann}, \citenamefont {Abrosimov}, \citenamefont {Becker}, \citenamefont
  {Pohl}, \citenamefont {Schenkel}, \citenamefont {Thewalt}, \citenamefont
  {Itoh} \emph {et~al.}}]{tyryshkin2012electron}%
  \BibitemOpen
  \bibfield  {author} {\bibinfo {author} {\bibfnamefont {A.~M.}\ \bibnamefont
  {Tyryshkin}}, \bibinfo {author} {\bibfnamefont {S.}~\bibnamefont {Tojo}},
  \bibinfo {author} {\bibfnamefont {J.~J.}\ \bibnamefont {Morton}}, \bibinfo
  {author} {\bibfnamefont {H.}~\bibnamefont {Riemann}}, \bibinfo {author}
  {\bibfnamefont {N.~V.}\ \bibnamefont {Abrosimov}}, \bibinfo {author}
  {\bibfnamefont {P.}~\bibnamefont {Becker}}, \bibinfo {author} {\bibfnamefont
  {H.-J.}\ \bibnamefont {Pohl}}, \bibinfo {author} {\bibfnamefont
  {T.}~\bibnamefont {Schenkel}}, \bibinfo {author} {\bibfnamefont {M.~L.}\
  \bibnamefont {Thewalt}}, \bibinfo {author} {\bibfnamefont {K.~M.}\
  \bibnamefont {Itoh}},  \emph {et~al.},\ }\href@noop {} {\bibfield  {journal}
  {\bibinfo  {journal} {Nature materials}\ }\textbf {\bibinfo {volume} {11}},\
  \bibinfo {pages} {143} (\bibinfo {year} {2012})}\BibitemShut {NoStop}%
\bibitem [{\citenamefont {Balasubramanian}\ \emph {et~al.}(2009)\citenamefont
  {Balasubramanian}, \citenamefont {Neumann}, \citenamefont {Twitchen},
  \citenamefont {Markham}, \citenamefont {Kolesov}, \citenamefont {Mizuochi},
  \citenamefont {Isoya}, \citenamefont {Achard}, \citenamefont {Beck},
  \citenamefont {Tissler}, \citenamefont {Jacques}, \citenamefont {Hemmer},
  \citenamefont {Jelezko},\ and\ \citenamefont {Wrachtrup}}]{2009_NV_isotope}%
  \BibitemOpen
  \bibfield  {author} {\bibinfo {author} {\bibfnamefont {G.}~\bibnamefont
  {Balasubramanian}}, \bibinfo {author} {\bibfnamefont {P.}~\bibnamefont
  {Neumann}}, \bibinfo {author} {\bibfnamefont {D.}~\bibnamefont {Twitchen}},
  \bibinfo {author} {\bibfnamefont {M.}~\bibnamefont {Markham}}, \bibinfo
  {author} {\bibfnamefont {R.}~\bibnamefont {Kolesov}}, \bibinfo {author}
  {\bibfnamefont {N.}~\bibnamefont {Mizuochi}}, \bibinfo {author}
  {\bibfnamefont {J.}~\bibnamefont {Isoya}}, \bibinfo {author} {\bibfnamefont
  {J.}~\bibnamefont {Achard}}, \bibinfo {author} {\bibfnamefont
  {J.}~\bibnamefont {Beck}}, \bibinfo {author} {\bibfnamefont {J.}~\bibnamefont
  {Tissler}}, \bibinfo {author} {\bibfnamefont {V.}~\bibnamefont {Jacques}},
  \bibinfo {author} {\bibfnamefont {P.~R.}\ \bibnamefont {Hemmer}}, \bibinfo
  {author} {\bibfnamefont {F.}~\bibnamefont {Jelezko}}, \ and\ \bibinfo
  {author} {\bibfnamefont {J.}~\bibnamefont {Wrachtrup}},\ }\href@noop {}
  {\bibfield  {journal} {\bibinfo  {journal} {Nature Materials}\ }\textbf
  {\bibinfo {volume} {8}},\ \bibinfo {pages} {383} (\bibinfo {year}
  {2009})}\BibitemShut {NoStop}%
\bibitem [{\citenamefont {Robledo}\ \emph {et~al.}(2011)\citenamefont
  {Robledo}, \citenamefont {Childress}, \citenamefont {Bernien}, \citenamefont
  {Hensen}, \citenamefont {Alkemade},\ and\ \citenamefont
  {Hanson}}]{robledo2011high}%
  \BibitemOpen
  \bibfield  {author} {\bibinfo {author} {\bibfnamefont {L.}~\bibnamefont
  {Robledo}}, \bibinfo {author} {\bibfnamefont {L.}~\bibnamefont {Childress}},
  \bibinfo {author} {\bibfnamefont {H.}~\bibnamefont {Bernien}}, \bibinfo
  {author} {\bibfnamefont {B.}~\bibnamefont {Hensen}}, \bibinfo {author}
  {\bibfnamefont {P.~F.}\ \bibnamefont {Alkemade}}, \ and\ \bibinfo {author}
  {\bibfnamefont {R.}~\bibnamefont {Hanson}},\ }\href@noop {} {\bibfield
  {journal} {\bibinfo  {journal} {Nature}\ }\textbf {\bibinfo {volume} {477}},\
  \bibinfo {pages} {574} (\bibinfo {year} {2011})}\BibitemShut {NoStop}%
\bibitem [{\citenamefont {Childress}\ and\ \citenamefont
  {Hanson}(2013)}]{childress2013diamond}%
  \BibitemOpen
  \bibfield  {author} {\bibinfo {author} {\bibfnamefont {L.}~\bibnamefont
  {Childress}}\ and\ \bibinfo {author} {\bibfnamefont {R.}~\bibnamefont
  {Hanson}},\ }\href@noop {} {\bibfield  {journal} {\bibinfo  {journal} {MRS
  bulletin}\ }\textbf {\bibinfo {volume} {38}},\ \bibinfo {pages} {134}
  (\bibinfo {year} {2013})}\BibitemShut {NoStop}%
\bibitem [{\citenamefont {Cappellaro}\ \emph {et~al.}(2009)\citenamefont
  {Cappellaro}, \citenamefont {Jiang}, \citenamefont {Hodges},\ and\
  \citenamefont {Lukin}}]{PhysRevLett.102.210502}%
  \BibitemOpen
  \bibfield  {author} {\bibinfo {author} {\bibfnamefont {P.}~\bibnamefont
  {Cappellaro}}, \bibinfo {author} {\bibfnamefont {L.}~\bibnamefont {Jiang}},
  \bibinfo {author} {\bibfnamefont {J.~S.}\ \bibnamefont {Hodges}}, \ and\
  \bibinfo {author} {\bibfnamefont {M.~D.}\ \bibnamefont {Lukin}},\ }\href
  {\doibase 10.1103/PhysRevLett.102.210502} {\bibfield  {journal} {\bibinfo
  {journal} {Phys. Rev. Lett.}\ }\textbf {\bibinfo {volume} {102}},\ \bibinfo
  {pages} {210502} (\bibinfo {year} {2009})}\BibitemShut {NoStop}%
\bibitem [{\citenamefont {Acosta}\ \emph {et~al.}(2009)\citenamefont {Acosta},
  \citenamefont {Bauch}, \citenamefont {Ledbetter}, \citenamefont {Santori},
  \citenamefont {Fu}, \citenamefont {Barclay}, \citenamefont {Beausoleil},
  \citenamefont {Linget}, \citenamefont {Roch}, \citenamefont {Treussart},
  \citenamefont {Chemerisov}, \citenamefont {Gawlik},\ and\ \citenamefont
  {Budker}}]{PhysRevB.80.115202}%
  \BibitemOpen
  \bibfield  {author} {\bibinfo {author} {\bibfnamefont {V.~M.}\ \bibnamefont
  {Acosta}}, \bibinfo {author} {\bibfnamefont {E.}~\bibnamefont {Bauch}},
  \bibinfo {author} {\bibfnamefont {M.~P.}\ \bibnamefont {Ledbetter}}, \bibinfo
  {author} {\bibfnamefont {C.}~\bibnamefont {Santori}}, \bibinfo {author}
  {\bibfnamefont {K.-M.~C.}\ \bibnamefont {Fu}}, \bibinfo {author}
  {\bibfnamefont {P.~E.}\ \bibnamefont {Barclay}}, \bibinfo {author}
  {\bibfnamefont {R.~G.}\ \bibnamefont {Beausoleil}}, \bibinfo {author}
  {\bibfnamefont {H.}~\bibnamefont {Linget}}, \bibinfo {author} {\bibfnamefont
  {J.~F.}\ \bibnamefont {Roch}}, \bibinfo {author} {\bibfnamefont
  {F.}~\bibnamefont {Treussart}}, \bibinfo {author} {\bibfnamefont
  {S.}~\bibnamefont {Chemerisov}}, \bibinfo {author} {\bibfnamefont
  {W.}~\bibnamefont {Gawlik}}, \ and\ \bibinfo {author} {\bibfnamefont
  {D.}~\bibnamefont {Budker}},\ }\href {\doibase 10.1103/PhysRevB.80.115202}
  {\bibfield  {journal} {\bibinfo  {journal} {Phys. Rev. B}\ }\textbf {\bibinfo
  {volume} {80}},\ \bibinfo {pages} {115202} (\bibinfo {year}
  {2009})}\BibitemShut {NoStop}%
\bibitem [{\citenamefont {Taylor}\ \emph {et~al.}(2008)\citenamefont {Taylor},
  \citenamefont {Cappellaro}, \citenamefont {Childress}, \citenamefont {Jiang},
  \citenamefont {Budker}, \citenamefont {Hemmer}, \citenamefont {Yacoby},
  \citenamefont {Walsworth},\ and\ \citenamefont {Lukin}}]{taylor2008high}%
  \BibitemOpen
  \bibfield  {author} {\bibinfo {author} {\bibfnamefont {J.}~\bibnamefont
  {Taylor}}, \bibinfo {author} {\bibfnamefont {P.}~\bibnamefont {Cappellaro}},
  \bibinfo {author} {\bibfnamefont {L.}~\bibnamefont {Childress}}, \bibinfo
  {author} {\bibfnamefont {L.}~\bibnamefont {Jiang}}, \bibinfo {author}
  {\bibfnamefont {D.}~\bibnamefont {Budker}}, \bibinfo {author} {\bibfnamefont
  {P.}~\bibnamefont {Hemmer}}, \bibinfo {author} {\bibfnamefont
  {A.}~\bibnamefont {Yacoby}}, \bibinfo {author} {\bibfnamefont
  {R.}~\bibnamefont {Walsworth}}, \ and\ \bibinfo {author} {\bibfnamefont
  {M.}~\bibnamefont {Lukin}},\ }\href@noop {} {\bibfield  {journal} {\bibinfo
  {journal} {Nature Physics}\ }\textbf {\bibinfo {volume} {4}},\ \bibinfo
  {pages} {810} (\bibinfo {year} {2008})}\BibitemShut {NoStop}%
\bibitem [{\citenamefont {Bauch}\ \emph {et~al.}(2018)\citenamefont {Bauch},
  \citenamefont {Hart}, \citenamefont {Schloss}, \citenamefont {Turner},
  \citenamefont {Barry}, \citenamefont {Kehayias}, \citenamefont {Singh},\ and\
  \citenamefont {Walsworth}}]{bauch2018ultralong}%
  \BibitemOpen
  \bibfield  {author} {\bibinfo {author} {\bibfnamefont {E.}~\bibnamefont
  {Bauch}}, \bibinfo {author} {\bibfnamefont {C.~A.}\ \bibnamefont {Hart}},
  \bibinfo {author} {\bibfnamefont {J.~M.}\ \bibnamefont {Schloss}}, \bibinfo
  {author} {\bibfnamefont {M.~J.}\ \bibnamefont {Turner}}, \bibinfo {author}
  {\bibfnamefont {J.~F.}\ \bibnamefont {Barry}}, \bibinfo {author}
  {\bibfnamefont {P.}~\bibnamefont {Kehayias}}, \bibinfo {author}
  {\bibfnamefont {S.}~\bibnamefont {Singh}}, \ and\ \bibinfo {author}
  {\bibfnamefont {R.~L.}\ \bibnamefont {Walsworth}},\ }\href@noop {} {\bibfield
   {journal} {\bibinfo  {journal} {Physical Review X}\ }\textbf {\bibinfo
  {volume} {8}},\ \bibinfo {pages} {031025} (\bibinfo {year}
  {2018})}\BibitemShut {NoStop}%
\bibitem [{\citenamefont {Ajoy}\ \emph {et~al.}(2015)\citenamefont {Ajoy},
  \citenamefont {Bissbort}, \citenamefont {Lukin}, \citenamefont {Walsworth},\
  and\ \citenamefont {Cappellaro}}]{PhysRevX.5.011001}%
  \BibitemOpen
  \bibfield  {author} {\bibinfo {author} {\bibfnamefont {A.}~\bibnamefont
  {Ajoy}}, \bibinfo {author} {\bibfnamefont {U.}~\bibnamefont {Bissbort}},
  \bibinfo {author} {\bibfnamefont {M.~D.}\ \bibnamefont {Lukin}}, \bibinfo
  {author} {\bibfnamefont {R.~L.}\ \bibnamefont {Walsworth}}, \ and\ \bibinfo
  {author} {\bibfnamefont {P.}~\bibnamefont {Cappellaro}},\ }\href {\doibase
  10.1103/PhysRevX.5.011001} {\bibfield  {journal} {\bibinfo  {journal} {Phys.
  Rev. X}\ }\textbf {\bibinfo {volume} {5}},\ \bibinfo {pages} {011001}
  (\bibinfo {year} {2015})}\BibitemShut {NoStop}%
\bibitem [{\citenamefont {Staudacher}\ \emph {et~al.}(2013)\citenamefont
  {Staudacher}, \citenamefont {Shi}, \citenamefont {Pezzagna}, \citenamefont
  {Meijer}, \citenamefont {Du}, \citenamefont {Meriles}, \citenamefont
  {Reinhard},\ and\ \citenamefont {Wrachtrup}}]{staudacher2013nuclear}%
  \BibitemOpen
  \bibfield  {author} {\bibinfo {author} {\bibfnamefont {T.}~\bibnamefont
  {Staudacher}}, \bibinfo {author} {\bibfnamefont {F.}~\bibnamefont {Shi}},
  \bibinfo {author} {\bibfnamefont {S.}~\bibnamefont {Pezzagna}}, \bibinfo
  {author} {\bibfnamefont {J.}~\bibnamefont {Meijer}}, \bibinfo {author}
  {\bibfnamefont {J.}~\bibnamefont {Du}}, \bibinfo {author} {\bibfnamefont
  {C.}~\bibnamefont {Meriles}}, \bibinfo {author} {\bibfnamefont
  {F.}~\bibnamefont {Reinhard}}, \ and\ \bibinfo {author} {\bibfnamefont
  {J.}~\bibnamefont {Wrachtrup}},\ }\href@noop {} {\bibfield  {journal}
  {\bibinfo  {journal} {Science}\ }\textbf {\bibinfo {volume} {339}},\ \bibinfo
  {pages} {561} (\bibinfo {year} {2013})}\BibitemShut {NoStop}%
\bibitem [{\citenamefont {Mamin}\ \emph {et~al.}(2013)\citenamefont {Mamin},
  \citenamefont {Kim}, \citenamefont {Sherwood}, \citenamefont {Rettner},
  \citenamefont {Ohno}, \citenamefont {Awschalom},\ and\ \citenamefont
  {Rugar}}]{mamin2013nanoscale}%
  \BibitemOpen
  \bibfield  {author} {\bibinfo {author} {\bibfnamefont {H.}~\bibnamefont
  {Mamin}}, \bibinfo {author} {\bibfnamefont {M.}~\bibnamefont {Kim}}, \bibinfo
  {author} {\bibfnamefont {M.}~\bibnamefont {Sherwood}}, \bibinfo {author}
  {\bibfnamefont {C.}~\bibnamefont {Rettner}}, \bibinfo {author} {\bibfnamefont
  {K.}~\bibnamefont {Ohno}}, \bibinfo {author} {\bibfnamefont {D.}~\bibnamefont
  {Awschalom}}, \ and\ \bibinfo {author} {\bibfnamefont {D.}~\bibnamefont
  {Rugar}},\ }\href@noop {} {\bibfield  {journal} {\bibinfo  {journal}
  {Science}\ }\textbf {\bibinfo {volume} {339}},\ \bibinfo {pages} {557}
  (\bibinfo {year} {2013})}\BibitemShut {NoStop}%
\bibitem [{\citenamefont {Sushkov}\ \emph {et~al.}(2014)\citenamefont
  {Sushkov}, \citenamefont {Lovchinsky}, \citenamefont {Chisholm},
  \citenamefont {Walsworth}, \citenamefont {Park},\ and\ \citenamefont
  {Lukin}}]{sushkov2014magnetic}%
  \BibitemOpen
  \bibfield  {author} {\bibinfo {author} {\bibfnamefont {A.}~\bibnamefont
  {Sushkov}}, \bibinfo {author} {\bibfnamefont {I.}~\bibnamefont {Lovchinsky}},
  \bibinfo {author} {\bibfnamefont {N.}~\bibnamefont {Chisholm}}, \bibinfo
  {author} {\bibfnamefont {R.}~\bibnamefont {Walsworth}}, \bibinfo {author}
  {\bibfnamefont {H.}~\bibnamefont {Park}}, \ and\ \bibinfo {author}
  {\bibfnamefont {M.}~\bibnamefont {Lukin}},\ }\href@noop {} {\bibfield
  {journal} {\bibinfo  {journal} {Physical review letters}\ }\textbf {\bibinfo
  {volume} {113}},\ \bibinfo {pages} {197601} (\bibinfo {year}
  {2014})}\BibitemShut {NoStop}%
\bibitem [{\citenamefont {Arndt}\ \emph {et~al.}(1993)\citenamefont {Arndt},
  \citenamefont {Kanorsky}, \citenamefont {Weis},\ and\ \citenamefont
  {H{\"a}nsch}}]{arndt1993can}%
  \BibitemOpen
  \bibfield  {author} {\bibinfo {author} {\bibfnamefont {M.}~\bibnamefont
  {Arndt}}, \bibinfo {author} {\bibfnamefont {S.}~\bibnamefont {Kanorsky}},
  \bibinfo {author} {\bibfnamefont {A.}~\bibnamefont {Weis}}, \ and\ \bibinfo
  {author} {\bibfnamefont {T.}~\bibnamefont {H{\"a}nsch}},\ }\href@noop {}
  {\bibfield  {journal} {\bibinfo  {journal} {Physics Letters A}\ }\textbf
  {\bibinfo {volume} {174}},\ \bibinfo {pages} {298} (\bibinfo {year}
  {1993})}\BibitemShut {NoStop}%
\bibitem [{\citenamefont {Kinoshita}\ \emph {et~al.}(1994)\citenamefont
  {Kinoshita}, \citenamefont {Takahashi},\ and\ \citenamefont
  {Yabuzaki}}]{kinoshita1994optical}%
  \BibitemOpen
  \bibfield  {author} {\bibinfo {author} {\bibfnamefont {T.}~\bibnamefont
  {Kinoshita}}, \bibinfo {author} {\bibfnamefont {Y.}~\bibnamefont
  {Takahashi}}, \ and\ \bibinfo {author} {\bibfnamefont {T.}~\bibnamefont
  {Yabuzaki}},\ }\href@noop {} {\bibfield  {journal} {\bibinfo  {journal}
  {Physical Review B}\ }\textbf {\bibinfo {volume} {49}},\ \bibinfo {pages}
  {3648} (\bibinfo {year} {1994})}\BibitemShut {NoStop}%
\bibitem [{\citenamefont {Vutha}\ \emph
  {et~al.}(2018{\natexlab{a}})\citenamefont {Vutha}, \citenamefont
  {Horbatsch},\ and\ \citenamefont {Hessels}}]{vutha2018oriented}%
  \BibitemOpen
  \bibfield  {author} {\bibinfo {author} {\bibfnamefont {A.}~\bibnamefont
  {Vutha}}, \bibinfo {author} {\bibfnamefont {M.}~\bibnamefont {Horbatsch}}, \
  and\ \bibinfo {author} {\bibfnamefont {E.}~\bibnamefont {Hessels}},\
  }\href@noop {} {\bibfield  {journal} {\bibinfo  {journal} {Atoms}\ }\textbf
  {\bibinfo {volume} {6}},\ \bibinfo {pages} {3} (\bibinfo {year}
  {2018}{\natexlab{a}})}\BibitemShut {NoStop}%
\bibitem [{\citenamefont {Vutha}\ \emph
  {et~al.}(2018{\natexlab{b}})\citenamefont {Vutha}, \citenamefont
  {Horbatsch},\ and\ \citenamefont {Hessels}}]{PhysRevA.98.032513}%
  \BibitemOpen
  \bibfield  {author} {\bibinfo {author} {\bibfnamefont {A.~C.}\ \bibnamefont
  {Vutha}}, \bibinfo {author} {\bibfnamefont {M.}~\bibnamefont {Horbatsch}}, \
  and\ \bibinfo {author} {\bibfnamefont {E.~A.}\ \bibnamefont {Hessels}},\
  }\href {\doibase 10.1103/PhysRevA.98.032513} {\bibfield  {journal} {\bibinfo
  {journal} {Phys. Rev. A}\ }\textbf {\bibinfo {volume} {98}},\ \bibinfo
  {pages} {032513} (\bibinfo {year} {2018}{\natexlab{b}})}\BibitemShut
  {NoStop}%
\bibitem [{\citenamefont {Lang}\ \emph {et~al.}(1999)\citenamefont {Lang},
  \citenamefont {Kanorsky}, \citenamefont {Eichler}, \citenamefont
  {M\"uller-Siebert}, \citenamefont {H\"ansch},\ and\ \citenamefont
  {Weis}}]{PhysRevA.60.3867}%
  \BibitemOpen
  \bibfield  {author} {\bibinfo {author} {\bibfnamefont {S.}~\bibnamefont
  {Lang}}, \bibinfo {author} {\bibfnamefont {S.}~\bibnamefont {Kanorsky}},
  \bibinfo {author} {\bibfnamefont {T.}~\bibnamefont {Eichler}}, \bibinfo
  {author} {\bibfnamefont {R.}~\bibnamefont {M\"uller-Siebert}}, \bibinfo
  {author} {\bibfnamefont {T.~W.}\ \bibnamefont {H\"ansch}}, \ and\ \bibinfo
  {author} {\bibfnamefont {A.}~\bibnamefont {Weis}},\ }\href {\doibase
  10.1103/PhysRevA.60.3867} {\bibfield  {journal} {\bibinfo  {journal} {Phys.
  Rev. A}\ }\textbf {\bibinfo {volume} {60}},\ \bibinfo {pages} {3867}
  (\bibinfo {year} {1999})}\BibitemShut {NoStop}%
\bibitem [{\citenamefont {Kanorsky}\ \emph {et~al.}(1996)\citenamefont
  {Kanorsky}, \citenamefont {Lang}, \citenamefont {L\"ucke}, \citenamefont
  {Ross}, \citenamefont {H\"ansch},\ and\ \citenamefont {Weis}}]{Weis1996}%
  \BibitemOpen
  \bibfield  {author} {\bibinfo {author} {\bibfnamefont {S.~I.}\ \bibnamefont
  {Kanorsky}}, \bibinfo {author} {\bibfnamefont {S.}~\bibnamefont {Lang}},
  \bibinfo {author} {\bibfnamefont {S.}~\bibnamefont {L\"ucke}}, \bibinfo
  {author} {\bibfnamefont {S.~B.}\ \bibnamefont {Ross}}, \bibinfo {author}
  {\bibfnamefont {T.~W.}\ \bibnamefont {H\"ansch}}, \ and\ \bibinfo {author}
  {\bibfnamefont {A.}~\bibnamefont {Weis}},\ }\href {\doibase
  10.1103/PhysRevA.54.R1010} {\bibfield  {journal} {\bibinfo  {journal} {Phys.
  Rev. A}\ }\textbf {\bibinfo {volume} {54}},\ \bibinfo {pages} {R1010}
  (\bibinfo {year} {1996})}\BibitemShut {NoStop}%
\bibitem [{\citenamefont {Moroshkin}\ \emph {et~al.}(2006)\citenamefont
  {Moroshkin}, \citenamefont {Hofer}, \citenamefont {Ulzega},\ and\
  \citenamefont {Weis}}]{moroshkin2006spectroscopy}%
  \BibitemOpen
  \bibfield  {author} {\bibinfo {author} {\bibfnamefont {P.}~\bibnamefont
  {Moroshkin}}, \bibinfo {author} {\bibfnamefont {A.}~\bibnamefont {Hofer}},
  \bibinfo {author} {\bibfnamefont {S.}~\bibnamefont {Ulzega}}, \ and\ \bibinfo
  {author} {\bibfnamefont {A.}~\bibnamefont {Weis}},\ }\href@noop {} {\bibfield
   {journal} {\bibinfo  {journal} {Low Temperature Physics}\ }\textbf {\bibinfo
  {volume} {32}},\ \bibinfo {pages} {981} (\bibinfo {year} {2006})}\BibitemShut
  {NoStop}%
\bibitem [{\citenamefont {Weyhmann}\ and\ \citenamefont
  {Pipkin}(1965)}]{PhysRev.137.A490}%
  \BibitemOpen
  \bibfield  {author} {\bibinfo {author} {\bibfnamefont {W.}~\bibnamefont
  {Weyhmann}}\ and\ \bibinfo {author} {\bibfnamefont {F.~M.}\ \bibnamefont
  {Pipkin}},\ }\href {\doibase 10.1103/PhysRev.137.A490} {\bibfield  {journal}
  {\bibinfo  {journal} {Phys. Rev.}\ }\textbf {\bibinfo {volume} {137}},\
  \bibinfo {pages} {A490} (\bibinfo {year} {1965})}\BibitemShut {NoStop}%
\bibitem [{\citenamefont {Xu}\ \emph {et~al.}(2011)\citenamefont {Xu},
  \citenamefont {Hu}, \citenamefont {Singh}, \citenamefont {Bailey},
  \citenamefont {Lu}, \citenamefont {Mueller}, \citenamefont {O'Connor},\ and\
  \citenamefont {Welp}}]{PhysRevLett.107.093001}%
  \BibitemOpen
  \bibfield  {author} {\bibinfo {author} {\bibfnamefont {C.-Y.}\ \bibnamefont
  {Xu}}, \bibinfo {author} {\bibfnamefont {S.-M.}\ \bibnamefont {Hu}}, \bibinfo
  {author} {\bibfnamefont {J.}~\bibnamefont {Singh}}, \bibinfo {author}
  {\bibfnamefont {K.}~\bibnamefont {Bailey}}, \bibinfo {author} {\bibfnamefont
  {Z.-T.}\ \bibnamefont {Lu}}, \bibinfo {author} {\bibfnamefont
  {P.}~\bibnamefont {Mueller}}, \bibinfo {author} {\bibfnamefont {T.~P.}\
  \bibnamefont {O'Connor}}, \ and\ \bibinfo {author} {\bibfnamefont
  {U.}~\bibnamefont {Welp}},\ }\href {\doibase 10.1103/PhysRevLett.107.093001}
  {\bibfield  {journal} {\bibinfo  {journal} {Phys. Rev. Lett.}\ }\textbf
  {\bibinfo {volume} {107}},\ \bibinfo {pages} {093001} (\bibinfo {year}
  {2011})}\BibitemShut {NoStop}%
\bibitem [{\citenamefont {Gaire}\ \emph {et~al.}(2019)\citenamefont {Gaire},
  \citenamefont {Raman},\ and\ \citenamefont {Parker}}]{PhysRevA.99.022505}%
  \BibitemOpen
  \bibfield  {author} {\bibinfo {author} {\bibfnamefont {V.}~\bibnamefont
  {Gaire}}, \bibinfo {author} {\bibfnamefont {C.~S.}\ \bibnamefont {Raman}}, \
  and\ \bibinfo {author} {\bibfnamefont {C.~V.}\ \bibnamefont {Parker}},\
  }\href {\doibase 10.1103/PhysRevA.99.022505} {\bibfield  {journal} {\bibinfo
  {journal} {Phys. Rev. A}\ }\textbf {\bibinfo {volume} {99}},\ \bibinfo
  {pages} {022505} (\bibinfo {year} {2019})}\BibitemShut {NoStop}%
\bibitem [{\citenamefont {Kupferman}\ and\ \citenamefont
  {Pipkin}(1968)}]{PhysRev.166.207}%
  \BibitemOpen
  \bibfield  {author} {\bibinfo {author} {\bibfnamefont {S.~L.}\ \bibnamefont
  {Kupferman}}\ and\ \bibinfo {author} {\bibfnamefont {F.~M.}\ \bibnamefont
  {Pipkin}},\ }\href {\doibase 10.1103/PhysRev.166.207} {\bibfield  {journal}
  {\bibinfo  {journal} {Phys. Rev.}\ }\textbf {\bibinfo {volume} {166}},\
  \bibinfo {pages} {207} (\bibinfo {year} {1968})}\BibitemShut {NoStop}%
\bibitem [{\citenamefont {Kanagin}\ \emph {et~al.}(2013)\citenamefont
  {Kanagin}, \citenamefont {Regmi}, \citenamefont {Pathak},\ and\ \citenamefont
  {Weinstein}}]{kanagin2013optical}%
  \BibitemOpen
  \bibfield  {author} {\bibinfo {author} {\bibfnamefont {A.~N.}\ \bibnamefont
  {Kanagin}}, \bibinfo {author} {\bibfnamefont {S.~K.}\ \bibnamefont {Regmi}},
  \bibinfo {author} {\bibfnamefont {P.}~\bibnamefont {Pathak}}, \ and\ \bibinfo
  {author} {\bibfnamefont {J.~D.}\ \bibnamefont {Weinstein}},\ }\href@noop {}
  {\bibfield  {journal} {\bibinfo  {journal} {Physical Review A}\ }\textbf
  {\bibinfo {volume} {88}},\ \bibinfo {pages} {063404} (\bibinfo {year}
  {2013})}\BibitemShut {NoStop}%
\bibitem [{\citenamefont {Momose}\ and\ \citenamefont
  {Shida}(1998)}]{Momose1998}%
  \BibitemOpen
  \bibfield  {author} {\bibinfo {author} {\bibfnamefont {T.}~\bibnamefont
  {Momose}}\ and\ \bibinfo {author} {\bibfnamefont {T.}~\bibnamefont {Shida}},\
  }\href@noop {} {\bibfield  {journal} {\bibinfo  {journal} {Bulletin of the
  Chemical Society of Japan}\ }\textbf {\bibinfo {volume} {71}},\ \bibinfo
  {pages} {1} (\bibinfo {year} {1998})}\BibitemShut {NoStop}%
\bibitem [{\citenamefont {Upadhyay}\ \emph {et~al.}(2016)\citenamefont
  {Upadhyay}, \citenamefont {Kanagin}, \citenamefont {Hartzell}, \citenamefont
  {Christy}, \citenamefont {Arnott}, \citenamefont {Momose}, \citenamefont
  {Patterson},\ and\ \citenamefont {Weinstein}}]{upadhyay2016longitudinal}%
  \BibitemOpen
  \bibfield  {author} {\bibinfo {author} {\bibfnamefont {S.}~\bibnamefont
  {Upadhyay}}, \bibinfo {author} {\bibfnamefont {A.~N.}\ \bibnamefont
  {Kanagin}}, \bibinfo {author} {\bibfnamefont {C.}~\bibnamefont {Hartzell}},
  \bibinfo {author} {\bibfnamefont {T.}~\bibnamefont {Christy}}, \bibinfo
  {author} {\bibfnamefont {W.~P.}\ \bibnamefont {Arnott}}, \bibinfo {author}
  {\bibfnamefont {T.}~\bibnamefont {Momose}}, \bibinfo {author} {\bibfnamefont
  {D.}~\bibnamefont {Patterson}}, \ and\ \bibinfo {author} {\bibfnamefont
  {J.~D.}\ \bibnamefont {Weinstein}},\ }\href@noop {} {\bibfield  {journal}
  {\bibinfo  {journal} {Physical Review Letters}\ }\textbf {\bibinfo {volume}
  {117}},\ \bibinfo {pages} {175301} (\bibinfo {year} {2016})}\BibitemShut
  {NoStop}%
\bibitem [{\citenamefont {Upadhyay}\ \emph {et~al.}(2019)\citenamefont
  {Upadhyay}, \citenamefont {Dargyte}, \citenamefont {Prater}, \citenamefont
  {Dergachev}, \citenamefont {Varganov}, \citenamefont {Tscherbul},
  \citenamefont {Patterson},\ and\ \citenamefont
  {Weinstein}}]{PhysRevB.100.024106}%
  \BibitemOpen
  \bibfield  {author} {\bibinfo {author} {\bibfnamefont {S.}~\bibnamefont
  {Upadhyay}}, \bibinfo {author} {\bibfnamefont {U.}~\bibnamefont {Dargyte}},
  \bibinfo {author} {\bibfnamefont {R.~P.}\ \bibnamefont {Prater}}, \bibinfo
  {author} {\bibfnamefont {V.~D.}\ \bibnamefont {Dergachev}}, \bibinfo {author}
  {\bibfnamefont {S.~A.}\ \bibnamefont {Varganov}}, \bibinfo {author}
  {\bibfnamefont {T.~V.}\ \bibnamefont {Tscherbul}}, \bibinfo {author}
  {\bibfnamefont {D.}~\bibnamefont {Patterson}}, \ and\ \bibinfo {author}
  {\bibfnamefont {J.~D.}\ \bibnamefont {Weinstein}},\ }\href {\doibase
  10.1103/PhysRevB.100.024106} {\bibfield  {journal} {\bibinfo  {journal}
  {Phys. Rev. B}\ }\textbf {\bibinfo {volume} {100}},\ \bibinfo {pages}
  {024106} (\bibinfo {year} {2019})}\BibitemShut {NoStop}%
\bibitem [{\citenamefont {Hartzell}(2014)}]{HartzellThesis}%
  \BibitemOpen
  \bibfield  {author} {\bibinfo {author} {\bibfnamefont {C.}~\bibnamefont
  {Hartzell}},\ }\emph {\bibinfo {title} {Matrix Isolation of Rubidium in a
  Solid Para-Hydrogen Substrate}},\ \href@noop {} {\bibinfo {type} {{B.S.
  Thesis}}},\ \bibinfo  {school} {University of Nevada, Reno} (\bibinfo {year}
  {2014})\BibitemShut {NoStop}%
\bibitem [{\citenamefont {Kanagin}(2015)}]{KanaginThesis}%
  \BibitemOpen
  \bibfield  {author} {\bibinfo {author} {\bibfnamefont {A.~N.}\ \bibnamefont
  {Kanagin}},\ }\emph {\bibinfo {title} {Creation and Analysis of Para-Hydrogen
  Crystals}},\ \href@noop {} {\bibinfo {type} {{B.S. Thesis}}},\ \bibinfo
  {school} {University of Nevada, Reno} (\bibinfo {year} {2015})\BibitemShut
  {NoStop}%
\bibitem [{\citenamefont {Fajardo}\ and\ \citenamefont
  {Tam}(1998)}]{jcp.108.4237}%
  \BibitemOpen
  \bibfield  {author} {\bibinfo {author} {\bibfnamefont {M.~E.}\ \bibnamefont
  {Fajardo}}\ and\ \bibinfo {author} {\bibfnamefont {S.}~\bibnamefont {Tam}},\
  }\href {\doibase 10.1063/1.475822} {\bibfield  {journal} {\bibinfo  {journal}
  {The Journal of Chemical Physics}\ }\textbf {\bibinfo {volume} {108}},\
  \bibinfo {pages} {4237} (\bibinfo {year} {1998})}\BibitemShut {NoStop}%
\bibitem [{\citenamefont {Sansonetti}\ \emph {et~al.}(2013)\citenamefont
  {Sansonetti}, \citenamefont {Martin},\ and\ \citenamefont
  {Young}}]{NISTAtomicBasic}%
  \BibitemOpen
  \bibfield  {author} {\bibinfo {author} {\bibfnamefont {J.~E.}\ \bibnamefont
  {Sansonetti}}, \bibinfo {author} {\bibfnamefont {W.~C.}\ \bibnamefont
  {Martin}}, \ and\ \bibinfo {author} {\bibfnamefont {S.~L.}\ \bibnamefont
  {Young}},\ }\href {\doibase 10.18434/T4FW23} {\emph {\bibinfo {title}
  {Handbook of Basic Atomic Spectroscopic Data (version 1.1.3)}}}\ (\bibinfo
  {publisher} {NIST},\ \bibinfo {year} {2013})\ \bibinfo {note}
  {http://physics.nist.gov/PhysRefData/Handbook/}\BibitemShut {NoStop}%
\bibitem [{\citenamefont {Gerhardt}\ \emph {et~al.}(2012)\citenamefont
  {Gerhardt}, \citenamefont {Sin},\ and\ \citenamefont
  {Momose}}]{gerhardt2012excitation}%
  \BibitemOpen
  \bibfield  {author} {\bibinfo {author} {\bibfnamefont {I.}~\bibnamefont
  {Gerhardt}}, \bibinfo {author} {\bibfnamefont {K.}~\bibnamefont {Sin}}, \
  and\ \bibinfo {author} {\bibfnamefont {T.}~\bibnamefont {Momose}},\
  }\href@noop {} {\bibfield  {journal} {\bibinfo  {journal} {The Journal of
  chemical physics}\ }\textbf {\bibinfo {volume} {137}},\ \bibinfo {pages}
  {014507} (\bibinfo {year} {2012})}\BibitemShut {NoStop}%
\bibitem [{\citenamefont {Takahashi}\ \emph {et~al.}(1993)\citenamefont
  {Takahashi}, \citenamefont {Sano}, \citenamefont {Kinoshita},\ and\
  \citenamefont {Yabuzaki}}]{PhysRevLett.71.1035}%
  \BibitemOpen
  \bibfield  {author} {\bibinfo {author} {\bibfnamefont {Y.}~\bibnamefont
  {Takahashi}}, \bibinfo {author} {\bibfnamefont {K.}~\bibnamefont {Sano}},
  \bibinfo {author} {\bibfnamefont {T.}~\bibnamefont {Kinoshita}}, \ and\
  \bibinfo {author} {\bibfnamefont {T.}~\bibnamefont {Yabuzaki}},\ }\href
  {\doibase 10.1103/PhysRevLett.71.1035} {\bibfield  {journal} {\bibinfo
  {journal} {Phys. Rev. Lett.}\ }\textbf {\bibinfo {volume} {71}},\ \bibinfo
  {pages} {1035} (\bibinfo {year} {1993})}\BibitemShut {NoStop}%
\bibitem [{\citenamefont {Happer}(1972)}]{Happer72OptPumpReview}%
  \BibitemOpen
  \bibfield  {author} {\bibinfo {author} {\bibfnamefont {W.}~\bibnamefont
  {Happer}},\ }\href@noop {} {\bibfield  {journal} {\bibinfo  {journal}
  {Reviews of Modern Physics}\ }\textbf {\bibinfo {volume} {44}},\ \bibinfo
  {pages} {169} (\bibinfo {year} {1972})}\BibitemShut {NoStop}%
\bibitem [{\citenamefont {Metcalf}\ and\ \citenamefont {Van~der
  Straten}(1999)}]{metcalf1999laser}%
  \BibitemOpen
  \bibfield  {author} {\bibinfo {author} {\bibfnamefont {H.}~\bibnamefont
  {Metcalf}}\ and\ \bibinfo {author} {\bibfnamefont {P.}~\bibnamefont {Van~der
  Straten}},\ }\href@noop {} {\emph {\bibinfo {title} {Laser cooling and
  trapping of atoms}}}\ (\bibinfo  {publisher} {Springer, New-York},\ \bibinfo
  {year} {1999})\BibitemShut {NoStop}%
\bibitem [{\citenamefont {Eichler}\ \emph {et~al.}(2002)\citenamefont
  {Eichler}, \citenamefont {M\"uller-Siebert}, \citenamefont {Nettels},
  \citenamefont {Kanorsky},\ and\ \citenamefont
  {Weis}}]{PhysRevLett.88.123002}%
  \BibitemOpen
  \bibfield  {author} {\bibinfo {author} {\bibfnamefont {T.}~\bibnamefont
  {Eichler}}, \bibinfo {author} {\bibfnamefont {R.}~\bibnamefont
  {M\"uller-Siebert}}, \bibinfo {author} {\bibfnamefont {D.}~\bibnamefont
  {Nettels}}, \bibinfo {author} {\bibfnamefont {S.}~\bibnamefont {Kanorsky}}, \
  and\ \bibinfo {author} {\bibfnamefont {A.}~\bibnamefont {Weis}},\ }\href
  {\doibase 10.1103/PhysRevLett.88.123002} {\bibfield  {journal} {\bibinfo
  {journal} {Phys. Rev. Lett.}\ }\textbf {\bibinfo {volume} {88}},\ \bibinfo
  {pages} {123002} (\bibinfo {year} {2002})}\BibitemShut {NoStop}%
\bibitem [{\citenamefont {Arimondo}\ \emph {et~al.}(1977)\citenamefont
  {Arimondo}, \citenamefont {Inguscio},\ and\ \citenamefont
  {Violino}}]{arimondo1977experimental}%
  \BibitemOpen
  \bibfield  {author} {\bibinfo {author} {\bibfnamefont {E.}~\bibnamefont
  {Arimondo}}, \bibinfo {author} {\bibfnamefont {M.}~\bibnamefont {Inguscio}},
  \ and\ \bibinfo {author} {\bibfnamefont {P.}~\bibnamefont {Violino}},\
  }\href@noop {} {\bibfield  {journal} {\bibinfo  {journal} {Reviews of Modern
  Physics}\ }\textbf {\bibinfo {volume} {49}},\ \bibinfo {pages} {31} (\bibinfo
  {year} {1977})}\BibitemShut {NoStop}%
\bibitem [{\citenamefont {Tiecke}(2010)}]{tiecke2010properties}%
  \BibitemOpen
  \bibfield  {author} {\bibinfo {author} {\bibfnamefont {T.}~\bibnamefont
  {Tiecke}},\ }\href@noop {} {\bibfield  {journal} {\bibinfo  {journal}
  {University of Amsterdam, The Netherlands, Thesis}\ ,\ \bibinfo {pages} {12}}
  (\bibinfo {year} {2010})}\BibitemShut {NoStop}%
\bibitem [{\citenamefont {Angel}\ and\ \citenamefont
  {Sandars}(1968)}]{angel1968hyperfine}%
  \BibitemOpen
  \bibfield  {author} {\bibinfo {author} {\bibfnamefont {J.~R.~P.}\
  \bibnamefont {Angel}}\ and\ \bibinfo {author} {\bibfnamefont
  {P.}~\bibnamefont {Sandars}},\ }\href@noop {} {\bibfield  {journal} {\bibinfo
   {journal} {Proceedings of the Royal Society of London. Series A.
  Mathematical and Physical Sciences}\ }\textbf {\bibinfo {volume} {305}},\
  \bibinfo {pages} {125} (\bibinfo {year} {1968})}\BibitemShut {NoStop}%
\bibitem [{\citenamefont {Dzuba}\ \emph {et~al.}(2010)\citenamefont {Dzuba},
  \citenamefont {Flambaum}, \citenamefont {Beloy},\ and\ \citenamefont
  {Derevianko}}]{dzuba2010hyperfine}%
  \BibitemOpen
  \bibfield  {author} {\bibinfo {author} {\bibfnamefont {V.}~\bibnamefont
  {Dzuba}}, \bibinfo {author} {\bibfnamefont {V.}~\bibnamefont {Flambaum}},
  \bibinfo {author} {\bibfnamefont {K.}~\bibnamefont {Beloy}}, \ and\ \bibinfo
  {author} {\bibfnamefont {A.}~\bibnamefont {Derevianko}},\ }\href@noop {}
  {\bibfield  {journal} {\bibinfo  {journal} {Physical Review A}\ }\textbf
  {\bibinfo {volume} {82}},\ \bibinfo {pages} {062513} (\bibinfo {year}
  {2010})}\BibitemShut {NoStop}%
\bibitem [{\citenamefont {Robyr}\ \emph {et~al.}(2014)\citenamefont {Robyr},
  \citenamefont {Knowles},\ and\ \citenamefont {Weis}}]{robyr2014measurement}%
  \BibitemOpen
  \bibfield  {author} {\bibinfo {author} {\bibfnamefont {J.-L.}\ \bibnamefont
  {Robyr}}, \bibinfo {author} {\bibfnamefont {P.}~\bibnamefont {Knowles}}, \
  and\ \bibinfo {author} {\bibfnamefont {A.}~\bibnamefont {Weis}},\ }\href@noop
  {} {\bibfield  {journal} {\bibinfo  {journal} {Physical Review A}\ }\textbf
  {\bibinfo {volume} {90}},\ \bibinfo {pages} {012505} (\bibinfo {year}
  {2014})}\BibitemShut {NoStop}%
\bibitem [{\citenamefont {Ulzega}\ \emph {et~al.}(2006)\citenamefont {Ulzega},
  \citenamefont {Hofer}, \citenamefont {Moroshkin},\ and\ \citenamefont
  {Weis}}]{Ulzega:06}%
  \BibitemOpen
  \bibfield  {author} {\bibinfo {author} {\bibfnamefont {S.}~\bibnamefont
  {Ulzega}}, \bibinfo {author} {\bibfnamefont {A.}~\bibnamefont {Hofer}},
  \bibinfo {author} {\bibfnamefont {P.}~\bibnamefont {Moroshkin}}, \ and\
  \bibinfo {author} {\bibfnamefont {A.}~\bibnamefont {Weis}},\ }\href@noop {}
  {\bibfield  {journal} {\bibinfo  {journal} {EPL (Europhysics Letters)}\
  }\textbf {\bibinfo {volume} {76}},\ \bibinfo {pages} {1074} (\bibinfo {year}
  {2006})}\BibitemShut {NoStop}%
\bibitem [{\citenamefont {Bloembergen}\ \emph {et~al.}(1948)\citenamefont
  {Bloembergen}, \citenamefont {Purcell},\ and\ \citenamefont
  {Pound}}]{PhysRev.73.679}%
  \BibitemOpen
  \bibfield  {author} {\bibinfo {author} {\bibfnamefont {N.}~\bibnamefont
  {Bloembergen}}, \bibinfo {author} {\bibfnamefont {E.~M.}\ \bibnamefont
  {Purcell}}, \ and\ \bibinfo {author} {\bibfnamefont {R.~V.}\ \bibnamefont
  {Pound}},\ }\href {\doibase 10.1103/PhysRev.73.679} {\bibfield  {journal}
  {\bibinfo  {journal} {Phys. Rev.}\ }\textbf {\bibinfo {volume} {73}},\
  \bibinfo {pages} {679} (\bibinfo {year} {1948})}\BibitemShut {NoStop}%
\bibitem [{\citenamefont {Murphy}(2009)}]{ESR1}%
  \BibitemOpen
  \bibfield  {author} {\bibinfo {author} {\bibfnamefont {D.~M.}\ \bibnamefont
  {Murphy}},\ }\href@noop {} {\emph {\bibinfo {title} {Electron Paramagnetic
  Resonance Spectroscopy of Polycrystalline Oxide Systems}}},\ edited by\
  \bibinfo {editor} {\bibfnamefont {S.~D.}\ \bibnamefont {Jackson}}\ and\
  \bibinfo {editor} {\bibfnamefont {J.~S.~J.}\ \bibnamefont {Hargreaves}}\
  (\bibinfo  {publisher} {WILEY-VCH},\ \bibinfo {year} {2009})\BibitemShut
  {NoStop}%
\bibitem [{\citenamefont {Carrington}\ and\ \citenamefont
  {McLachlan}(1967)}]{Carrington67}%
  \BibitemOpen
  \bibfield  {author} {\bibinfo {author} {\bibfnamefont {A.}~\bibnamefont
  {Carrington}}\ and\ \bibinfo {author} {\bibfnamefont {A.~D.}\ \bibnamefont
  {McLachlan}},\ }\href@noop {} {\emph {\bibinfo {title} {Introduction to
  Magnetic Resonance}}}\ (\bibinfo  {publisher} {Harper},\ \bibinfo {year}
  {1967})\BibitemShut {NoStop}%
\bibitem [{\citenamefont {Slichter}(1990)}]{slichter1990}%
  \BibitemOpen
  \bibfield  {author} {\bibinfo {author} {\bibfnamefont {C.~P.}\ \bibnamefont
  {Slichter}},\ }\href@noop {} {\emph {\bibinfo {title} {Principles of Magnetic
  Resonance}}},\ \bibinfo {edition} {3rd}\ ed.\ (\bibinfo  {publisher}
  {Springer},\ \bibinfo {year} {1990})\BibitemShut {NoStop}%
\bibitem [{\citenamefont {Lund}\ \emph {et~al.}(2011)\citenamefont {Lund},
  \citenamefont {Shiotani},\ and\ \citenamefont {Shimada}}]{Lund:11}%
  \BibitemOpen
  \bibfield  {author} {\bibinfo {author} {\bibfnamefont {A.}~\bibnamefont
  {Lund}}, \bibinfo {author} {\bibfnamefont {M.}~\bibnamefont {Shiotani}}, \
  and\ \bibinfo {author} {\bibfnamefont {S.}~\bibnamefont {Shimada}},\
  }\href@noop {} {\emph {\bibinfo {title} {Principles and applications of ESR
  spectroscopy}}}\ (\bibinfo  {publisher} {Springer Science \& Business
  Media},\ \bibinfo {year} {2011})\BibitemShut {NoStop}%
\bibitem [{\citenamefont {Tscherbul}\ \emph {et~al.}(2012)\citenamefont
  {Tscherbul}, \citenamefont {Grinev}, \citenamefont {Yu}, \citenamefont
  {Dalgarno}, \citenamefont {K{\l}os}, \citenamefont {Ma},\ and\ \citenamefont
  {Alexander}}]{Tscherbul:12}%
  \BibitemOpen
  \bibfield  {author} {\bibinfo {author} {\bibfnamefont {T.~V.}\ \bibnamefont
  {Tscherbul}}, \bibinfo {author} {\bibfnamefont {T.~A.}\ \bibnamefont
  {Grinev}}, \bibinfo {author} {\bibfnamefont {H.-G.}\ \bibnamefont {Yu}},
  \bibinfo {author} {\bibfnamefont {A.}~\bibnamefont {Dalgarno}}, \bibinfo
  {author} {\bibfnamefont {J.}~\bibnamefont {K{\l}os}}, \bibinfo {author}
  {\bibfnamefont {L.}~\bibnamefont {Ma}}, \ and\ \bibinfo {author}
  {\bibfnamefont {M.~H.}\ \bibnamefont {Alexander}},\ }\href@noop {} {\bibfield
   {journal} {\bibinfo  {journal} {J. Chem. Phys.}\ }\textbf {\bibinfo {volume}
  {137}},\ \bibinfo {pages} {104302} (\bibinfo {year} {2012})}\BibitemShut
  {NoStop}%
\bibitem [{\citenamefont {Zare}(1988)}]{Zare:88}%
  \BibitemOpen
  \bibfield  {author} {\bibinfo {author} {\bibfnamefont {R.~N.}\ \bibnamefont
  {Zare}},\ }\href@noop {} {\emph {\bibinfo {title} {Angular Momentum}}}\
  (\bibinfo  {publisher} {Wiley, New York},\ \bibinfo {year}
  {1988})\BibitemShut {NoStop}%
\bibitem [{\citenamefont {Scharf}\ \emph
  {et~al.}(1993{\natexlab{a}})\citenamefont {Scharf}, \citenamefont {Martyna},
  \citenamefont {Li}, \citenamefont {Voth},\ and\ \citenamefont
  {Klein}}]{scharf1993nature}%
  \BibitemOpen
  \bibfield  {author} {\bibinfo {author} {\bibfnamefont {D.}~\bibnamefont
  {Scharf}}, \bibinfo {author} {\bibfnamefont {G.~J.}\ \bibnamefont {Martyna}},
  \bibinfo {author} {\bibfnamefont {D.}~\bibnamefont {Li}}, \bibinfo {author}
  {\bibfnamefont {G.~A.}\ \bibnamefont {Voth}}, \ and\ \bibinfo {author}
  {\bibfnamefont {M.~L.}\ \bibnamefont {Klein}},\ }\href@noop {} {\bibfield
  {journal} {\bibinfo  {journal} {The Journal of chemical physics}\ }\textbf
  {\bibinfo {volume} {99}},\ \bibinfo {pages} {9013} (\bibinfo {year}
  {1993}{\natexlab{a}})}\BibitemShut {NoStop}%
\bibitem [{\citenamefont {Deegan}\ and\ \citenamefont
  {Knowles}(1994)}]{SevaAbinitio1}%
  \BibitemOpen
  \bibfield  {author} {\bibinfo {author} {\bibfnamefont {M.~J.~O.}\
  \bibnamefont {Deegan}}\ and\ \bibinfo {author} {\bibfnamefont {P.~J.}\
  \bibnamefont {Knowles}},\ }\href@noop {} {\bibfield  {journal} {\bibinfo
  {journal} {Chem. Phys. Lett.}\ }\textbf {\bibinfo {volume} {227}},\ \bibinfo
  {pages} {321} (\bibinfo {year} {1994})}\BibitemShut {NoStop}%
\bibitem [{\citenamefont {Werner}\ \emph {et~al.}(2011)\citenamefont {Werner},
  \citenamefont {Knowles}, \citenamefont {Knizia}, \citenamefont {Manby},\ and\
  \citenamefont {Sch\"utz}}]{SevaAbinitio2}%
  \BibitemOpen
  \bibfield  {author} {\bibinfo {author} {\bibfnamefont {H.-J.}\ \bibnamefont
  {Werner}}, \bibinfo {author} {\bibfnamefont {P.~J.}\ \bibnamefont {Knowles}},
  \bibinfo {author} {\bibfnamefont {G.}~\bibnamefont {Knizia}}, \bibinfo
  {author} {\bibfnamefont {F.~R.}\ \bibnamefont {Manby}}, \ and\ \bibinfo
  {author} {\bibfnamefont {M.}~\bibnamefont {Sch\"utz}},\ }\href@noop {}
  {\bibfield  {journal} {\bibinfo  {journal} {Wiley Interdiscip. Rev.: Comput.
  Mol. Sci.}\ }\textbf {\bibinfo {volume} {2}},\ \bibinfo {pages} {242}
  (\bibinfo {year} {2011})}\BibitemShut {NoStop}%
\bibitem [{\citenamefont {Kendall}\ \emph {et~al.}(1992)\citenamefont
  {Kendall}, \citenamefont {Dunning},\ and\ \citenamefont
  {Harrison}}]{SevaAbinitio3}%
  \BibitemOpen
  \bibfield  {author} {\bibinfo {author} {\bibfnamefont {R.~A.}\ \bibnamefont
  {Kendall}}, \bibinfo {author} {\bibfnamefont {T.~H.}\ \bibnamefont
  {Dunning}}, \ and\ \bibinfo {author} {\bibfnamefont {R.~J.}\ \bibnamefont
  {Harrison}},\ }\href@noop {} {\bibfield  {journal} {\bibinfo  {journal} {J.
  Chem. Phys.}\ }\textbf {\bibinfo {volume} {96}},\ \bibinfo {pages} {6796}
  (\bibinfo {year} {1992})}\BibitemShut {NoStop}%
\bibitem [{\citenamefont {Arruda}\ \emph {et~al.}(2009)\citenamefont {Arruda},
  \citenamefont {Neto},\ and\ \citenamefont {Jorge}}]{SevaAbinitio4}%
  \BibitemOpen
  \bibfield  {author} {\bibinfo {author} {\bibfnamefont {P.~M.}\ \bibnamefont
  {Arruda}}, \bibinfo {author} {\bibfnamefont {A.~C.}\ \bibnamefont {Neto}}, \
  and\ \bibinfo {author} {\bibfnamefont {F.~E.}\ \bibnamefont {Jorge}},\
  }\href@noop {} {\bibfield  {journal} {\bibinfo  {journal} {Int. J. Quantum
  Chem.}\ }\textbf {\bibinfo {volume} {109}},\ \bibinfo {pages} {1189}
  (\bibinfo {year} {2009})}\BibitemShut {NoStop}%
\bibitem [{\citenamefont {Lim}\ \emph {et~al.}(2005)\citenamefont {Lim},
  \citenamefont {Schwerdtfeger}, \citenamefont {Metz},\ and\ \citenamefont
  {Stoll}}]{SevaAbinitio5}%
  \BibitemOpen
  \bibfield  {author} {\bibinfo {author} {\bibfnamefont {I.~S.}\ \bibnamefont
  {Lim}}, \bibinfo {author} {\bibfnamefont {P.}~\bibnamefont {Schwerdtfeger}},
  \bibinfo {author} {\bibfnamefont {B.}~\bibnamefont {Metz}}, \ and\ \bibinfo
  {author} {\bibfnamefont {H.}~\bibnamefont {Stoll}},\ }\href@noop {}
  {\bibfield  {journal} {\bibinfo  {journal} {J. Chem. Phys.}\ }\textbf
  {\bibinfo {volume} {122}},\ \bibinfo {pages} {104103} (\bibinfo {year}
  {2005})}\BibitemShut {NoStop}%
\bibitem [{\citenamefont {Boys}\ and\ \citenamefont
  {Bernardi}(1970)}]{SevaAbinitio6}%
  \BibitemOpen
  \bibfield  {author} {\bibinfo {author} {\bibfnamefont {S.~F.}\ \bibnamefont
  {Boys}}\ and\ \bibinfo {author} {\bibfnamefont {F.}~\bibnamefont
  {Bernardi}},\ }\href@noop {} {\bibfield  {journal} {\bibinfo  {journal} {Mol.
  Phys.}\ }\textbf {\bibinfo {volume} {19}},\ \bibinfo {pages} {553} (\bibinfo
  {year} {1970})}\BibitemShut {NoStop}%
\bibitem [{\citenamefont {Li}\ \emph {et~al.}(2010)\citenamefont {Li},
  \citenamefont {Roy},\ and\ \citenamefont {Le~Roy}}]{li2010adiabatic}%
  \BibitemOpen
  \bibfield  {author} {\bibinfo {author} {\bibfnamefont {H.}~\bibnamefont
  {Li}}, \bibinfo {author} {\bibfnamefont {P.-N.}\ \bibnamefont {Roy}}, \ and\
  \bibinfo {author} {\bibfnamefont {R.~J.}\ \bibnamefont {Le~Roy}},\
  }\href@noop {} {\bibfield  {journal} {\bibinfo  {journal} {The Journal of
  chemical physics}\ }\textbf {\bibinfo {volume} {133}},\ \bibinfo {pages}
  {104305} (\bibinfo {year} {2010})}\BibitemShut {NoStop}%
\bibitem [{\citenamefont {{J. F. Stanton}}(2019)}]{SevaAbinitio8}%
  \BibitemOpen
  \bibfield  {author} {\bibinfo {author} {\bibnamefont {{J. F. Stanton}}},\
  }\href@noop {} {\bibfield  {journal} {\bibinfo  {journal} {{\it et al},
  CFOUR, http://www.cfour.de}\ } (\bibinfo {year} {2019})}\BibitemShut
  {NoStop}%
\bibitem [{\citenamefont {Weigend}\ \emph {et~al.}(2002)\citenamefont
  {Weigend}, \citenamefont {Kohn},\ and\ \citenamefont
  {Hattig}}]{SevaAbinitio9}%
  \BibitemOpen
  \bibfield  {author} {\bibinfo {author} {\bibfnamefont {F.}~\bibnamefont
  {Weigend}}, \bibinfo {author} {\bibfnamefont {A.}~\bibnamefont {Kohn}}, \
  and\ \bibinfo {author} {\bibfnamefont {C.}~\bibnamefont {Hattig}},\
  }\href@noop {} {\bibfield  {journal} {\bibinfo  {journal} {J. Chem. Phys.}\
  }\textbf {\bibinfo {volume} {116}},\ \bibinfo {pages} {3175} (\bibinfo {year}
  {2002})}\BibitemShut {NoStop}%
\bibitem [{\citenamefont {Roos}\ \emph {et~al.}(2004)\citenamefont {Roos},
  \citenamefont {Veryazov},\ and\ \citenamefont {Widmark}}]{SevaAbinitio10}%
  \BibitemOpen
  \bibfield  {author} {\bibinfo {author} {\bibfnamefont {B.~O.}\ \bibnamefont
  {Roos}}, \bibinfo {author} {\bibfnamefont {V.}~\bibnamefont {Veryazov}}, \
  and\ \bibinfo {author} {\bibfnamefont {P.-O.}\ \bibnamefont {Widmark}},\
  }\href@noop {} {\bibfield  {journal} {\bibinfo  {journal} {Theor. Chem.
  Acc.}\ }\textbf {\bibinfo {volume} {111}},\ \bibinfo {pages} {345} (\bibinfo
  {year} {2004})}\BibitemShut {NoStop}%
\bibitem [{\citenamefont {Tscherbul}\ \emph {et~al.}(2009)\citenamefont
  {Tscherbul}, \citenamefont {Zhang}, \citenamefont {Sadeghpour},\ and\
  \citenamefont {Dalgarno}}]{Tscherbul:09}%
  \BibitemOpen
  \bibfield  {author} {\bibinfo {author} {\bibfnamefont {T.~V.}\ \bibnamefont
  {Tscherbul}}, \bibinfo {author} {\bibfnamefont {P.}~\bibnamefont {Zhang}},
  \bibinfo {author} {\bibfnamefont {H.~R.}\ \bibnamefont {Sadeghpour}}, \ and\
  \bibinfo {author} {\bibfnamefont {A.}~\bibnamefont {Dalgarno}},\ }\href@noop
  {} {\bibfield  {journal} {\bibinfo  {journal} {Phys. Rev. A}\ }\textbf
  {\bibinfo {volume} {79}},\ \bibinfo {pages} {062707} (\bibinfo {year}
  {2009})}\BibitemShut {NoStop}%
\bibitem [{\citenamefont {Tscherbul}\ \emph {et~al.}(2011)\citenamefont
  {Tscherbul}, \citenamefont {Zhang}, \citenamefont {Sadeghpour},\ and\
  \citenamefont {Dalgarno}}]{Tscherbul:11}%
  \BibitemOpen
  \bibfield  {author} {\bibinfo {author} {\bibfnamefont {T.~V.}\ \bibnamefont
  {Tscherbul}}, \bibinfo {author} {\bibfnamefont {P.}~\bibnamefont {Zhang}},
  \bibinfo {author} {\bibfnamefont {H.~R.}\ \bibnamefont {Sadeghpour}}, \ and\
  \bibinfo {author} {\bibfnamefont {A.}~\bibnamefont {Dalgarno}},\ }\href@noop
  {} {\bibfield  {journal} {\bibinfo  {journal} {Phys. Rev. Lett.}\ }\textbf
  {\bibinfo {volume} {107}},\ \bibinfo {pages} {023204} (\bibinfo {year}
  {2011})}\BibitemShut {NoStop}%
\bibitem [{\citenamefont {Anderson}\ \emph {et~al.}(1960)\citenamefont
  {Anderson}, \citenamefont {Pipkin},\ and\ \citenamefont
  {Baird~Jr}}]{SevaAbinitio11}%
  \BibitemOpen
  \bibfield  {author} {\bibinfo {author} {\bibfnamefont {L.~W.}\ \bibnamefont
  {Anderson}}, \bibinfo {author} {\bibfnamefont {F.~M.}\ \bibnamefont
  {Pipkin}}, \ and\ \bibinfo {author} {\bibfnamefont {J.~C.}\ \bibnamefont
  {Baird~Jr}},\ }\href@noop {} {\bibfield  {journal} {\bibinfo  {journal}
  {Physical Review}\ }\textbf {\bibinfo {volume} {120}},\ \bibinfo {pages}
  {1279} (\bibinfo {year} {1960})}\BibitemShut {NoStop}%
\bibitem [{\citenamefont {Silvera}(1980)}]{RevModPhys.52.393}%
  \BibitemOpen
  \bibfield  {author} {\bibinfo {author} {\bibfnamefont {I.~F.}\ \bibnamefont
  {Silvera}},\ }\href {\doibase 10.1103/RevModPhys.52.393} {\bibfield
  {journal} {\bibinfo  {journal} {Rev. Mod. Phys.}\ }\textbf {\bibinfo {volume}
  {52}},\ \bibinfo {pages} {393} (\bibinfo {year} {1980})}\BibitemShut
  {NoStop}%
\bibitem [{\citenamefont {Scharf}\ \emph
  {et~al.}(1993{\natexlab{b}})\citenamefont {Scharf}, \citenamefont {Martyna},
  \citenamefont {Li}, \citenamefont {Voth},\ and\ \citenamefont
  {Klein}}]{scharf:9013}%
  \BibitemOpen
  \bibfield  {author} {\bibinfo {author} {\bibfnamefont {D.}~\bibnamefont
  {Scharf}}, \bibinfo {author} {\bibfnamefont {G.~J.}\ \bibnamefont {Martyna}},
  \bibinfo {author} {\bibfnamefont {D.}~\bibnamefont {Li}}, \bibinfo {author}
  {\bibfnamefont {G.~A.}\ \bibnamefont {Voth}}, \ and\ \bibinfo {author}
  {\bibfnamefont {M.~L.}\ \bibnamefont {Klein}},\ }\href {\doibase
  10.1063/1.465569} {\bibfield  {journal} {\bibinfo  {journal} {The Journal of
  Chemical Physics}\ }\textbf {\bibinfo {volume} {99}},\ \bibinfo {pages}
  {9013} (\bibinfo {year} {1993}{\natexlab{b}})}\BibitemShut {NoStop}%
\bibitem [{\citenamefont {Steck}(2013)}]{steck2013rubidium85}%
  \BibitemOpen
  \bibfield  {author} {\bibinfo {author} {\bibfnamefont {D.~A.}\ \bibnamefont
  {Steck}},\ }\href {http://steck.us/alkalidata} {\enquote {\bibinfo {title}
  {Rubidium 85 {D} line data},}\ } (\bibinfo {year} {2013})\BibitemShut
  {NoStop}%
\bibitem [{\citenamefont {Pathak}(2014)}]{pawanthesis}%
  \BibitemOpen
  \bibfield  {author} {\bibinfo {author} {\bibfnamefont {P.}~\bibnamefont
  {Pathak}},\ }\emph {\bibinfo {title} {Absorption Spectrum of Rubidium in a
  Solid Neon Matrix}},\ \href@noop {} {\bibinfo {type} {{M.S. Thesis}}},\
  \bibinfo  {school} {University of Nevada, Reno} (\bibinfo {year}
  {2014})\BibitemShut {NoStop}%
\end{thebibliography}%

\end{document}